\documentclass[journal]{IEEEtran}

	\usepackage{ifpdf}

\usepackage{cite}

\ifCLASSINFOpdf
	\usepackage[pdftex,hiresbb]{graphicx}
\else
  \usepackage[dvips]{graphicx}
\fi

\usepackage[cmex10]{amsmath}
\usepackage{microtype}	
\usepackage{url}

\usepackage{enumitem}
\usepackage{multirow}
\usepackage{bigstrut}
\usepackage{amssymb,amsfonts}
\usepackage{bm}
	\usepackage[pdftex]{xcolor}
\usepackage[normalem]{ulem}
\usepackage{xspace}
\usepackage{textcomp}
\usepackage{algorithm}
\usepackage{algpseudocode}
\algrenewcommand\alglinenumber[1]{\footnotesize #1:}
\usepackage{xpatch}
\usepackage[caption=false]{subfig}
\usepackage{placeins}

\makeatletter
\newcommand{\removelatexerror}{\let\@latex@error\@gobble}
\makeatother

\newcommand{\rbt}{\mathbf{r}}
\newcommand{\ubt}{\mathbf{u}}
\newcommand{\wbt}{\mathbf{w}}
\newcommand{\ybt}{\mathbf{y}}
\newcommand{\Hbt}{\mathbf{H}}
\newcommand{\Ibt}{\mathbf{I}}
\newcommand{\xbt}{\mathbf{x}}
\newcommand{\Xbt}{\mathbf{X}}
\newcommand{\zero}{\mathbf{0}}
\newcommand{\mubt}{\boldsymbol{\mu}}
\newcommand{\Sigmabt}{\boldsymbol{\Sigma}}
\newcommand{\lambdabt}{\boldsymbol{\lambda}}

\makeatletter
\xpatchcmd{\algorithmic}{\itemsep\z@}{\itemsep=0.6ex plus 0.6ex  minus 0.6ex}{}{}
\makeatother

\usepackage{lettrine}
\newcommand{\introinitialdrop}[2]{%
  \lettrine[lines=2,lhang=0.33,nindent=0em]{#1}{\textsc{#2}}%
}

\makeatletter
\newcommand\fs@betterruled{%
  \def\@fs@cfont{\bfseries}\let\@fs@capt\floatc@ruled
  \def\@fs@pre{\vspace*{5pt}\hrule height.8pt depth0pt \kern2pt}%
  \def\@fs@post{\kern2pt\hrule\relax}%
  \def\@fs@mid{\kern2pt\hrule\kern2pt}%
  \let\@fs@iftopcapt\iftrue}
\floatstyle{betterruled}
\restylefloat{algorithm}
\makeatother

\newif\ifcleanversion
\cleanversiontrue

\ifcleanversion
  \AtBeginDocument{\colorlet{red}{black}}
  \renewcommand{\marginpar}[1]{}
\fi

\begin{document}
\title{Hamiltonian Monte Carlo-Based Near-Optimal MIMO Signal Detection}

\author{Junichiro Hagiwara,~\IEEEmembership{Member,~IEEE,}
Toshihiko Nishimura,~\IEEEmembership{Member,~IEEE,}
Takanori Sato,~\IEEEmembership{Member,~IEEE,}
Yasutaka Ogawa,~\IEEEmembership{Life~Fellow,~IEEE,}
and Takeo Ohgane,~\IEEEmembership{Member,~IEEE}%
\thanks{Manuscript received November 16, 2025.
This paper was presented in part at the GLOBECOM 2023, Kuala Lumpur, Malaysia, 4--8 December 2023~\cite{Globecom} (DOI: 10.1109/GLOBECOM54140.2023.10437162).
}
\thanks{%
Junichiro Hagiwara was with the Faculty of Information Science and Technology, Hokkaido University, Sapporo 060-0814, Japan. He is currently with the Faculty of Social Informatics, Mukogawa Women's University, Nishinomiya 663-8558, Japan (e-mail: hagijyun@gmail.com).
}
\thanks{%
Toshihiko Nishimura, Takanori Sato, Yasutaka Ogawa, and Takeo Ohgane are with the Faculty of Information Science and Technology, Hokkaido University, Sapporo 060-0814, Japan (e-mail: nishim@ist.hokudai.ac.jp; tksato@ist.hokudai.ac.jp; ogawa@ist.hokudai.ac.jp; ohgane@ist.hokudai.ac.jp).
}
\thanks{%
Color versions of one or more of the figures in this paper are available online at http://ieeexplore.ieee.org.
}
}

\maketitle

\begin{abstract}
Multiple-input multiple-output (MIMO) technology is essential for the optimal functioning of next-generation wireless networks; however, enhancing its signal-detection performance for improved spectral efficiency is challenging.
Here, we propose an approach that transforms the discrete MIMO detection problem into a continuous problem while leveraging the efficient Hamiltonian Monte Carlo algorithm.
For this continuous framework, we employ a mixture of $\bm{t}$-distributions as the prior distribution.
To improve the performance in the coded case further, we treat the likelihood's temperature parameter as a random variable and address its optimization.
This treatment leads to the adoption of a horseshoe density for the likelihood.
Theoretical analysis and extensive simulations demonstrate that our method achieves near-optimal detection performance while maintaining polynomial computational complexity.
This MIMO detection technique can accelerate the development of 6G mobile communication systems.
\end{abstract}


\section{Introduction} \label{sec:Introduction}
\introinitialdrop{T}{he} widespread adoption of 5G networks underscores the pivotal role of wireless communications in contemporary infrastructure.
Ongoing 6G research aims to enhance user experience by improving the utilization efficiency of the radio spectrum.
The multiple-input multiple-output (MIMO) technology is crucial for the effective utilization of the radio spectrum in next-generation wireless networks.
By employing multiple antennas at both ends, MIMO systems significantly boost the efficiency of wireless transmission.
Although conventional MIMO systems have been widely implemented, recent studies have focused on massive MIMO systems~\cite{An_Overview_of_Massive_MIMO}.

High-accuracy signal detection in MIMO systems is crucial for enhancing their overall transmission efficiency.
Generally, the maximum likelihood detection algorithm delivers optimal performance; however, its exhaustive search approach and associated computational costs limit its practical application in MIMO systems~\cite{chockalingam2014large}.
Linear-detection algorithms, such as the minimum mean-square error (MMSE) algorithm, offer computational efficiency but underperform compared to optimal techniques~\cite{chockalingam2014large}.
The key challenge lies in enhancing the signal-detection accuracy without exceeding the practical computational limits.
A promising approach involves adopting a stochastic framework for MIMO signal detection.
This perspective enables the theoretical incorporation of uncertainty and the application of advanced probabilistic methods in signal processing.

Stochastic approaches, including Bayesian methods, have been applied to improve the MIMO signal detection~\cite{Fifty_Years_of_MIMO_Detection,Massive_MIMO_Detection_Techniques}.
Continuous prior distributions are generally considered unsuitable for MIMO signal detection, with the exception of~\cite{Langevin} and our previously proposed methods~\cite{Globecom,WPMC,MatsumuraRCS2020a,MatsumuraRCS2020b,AsumiRCS2021,AsumiRCS,AsumiRCS2023,KasugaGC2022}.
In the mixed Gibbs sampling (MGS) method~\cite{MGS}, numerous samples are employed to approximate the posterior distribution based on a discrete prior distribution and Gibbs sampling~\cite{Gibbs}, a Markov chain Monte Carlo (MCMC) technique.
To improve MCMC exploration with a discrete prior distribution, \cite{ExcitedMCMC} introduces the dynamic scaling of the likelihood's temperature parameter, which impacts the Gibbs sampling exploration.
Alternatively, \cite{GradientBasedMCMC,GradientBasedMH} suggest a Metropolis--Hastings (MH) algorithm utilizing the likelihood gradient information.
Meanwhile, the expectation propagation (EP) method~\cite{EP} employs a discrete prior distribution and applies the EP algorithm~\cite{EPMinka} to approximate the posterior distribution parameters.
\color{red}
\marginpar{{\footnotesize [R3.8]}}%
Furthermore, in recent years, numerous approaches leveraging deep learning have emerged, with model-based methods applying deep unfolding to existing detection algorithms becoming mainstream.
Specifically, various methods have been proposed, including DetNet~\cite{DetNet} based on projected gradient descent, OAMP-Net2~\cite{OAMPNet2} based on orthogonal approximate message passing (OAMP)~\cite{OAMP}, and deep-unfolded interleaved detection and decoding~\cite{DUIDD} based on MMSE-parallel interference cancellation, all of which assume discrete priors.
\color{black}
Most stochastic MIMO detection studies have adopted a discrete prior distribution, reflecting the discrete nature of the transmission symbols.
Consequently, numerous powerful algorithms designed for continuous problems are not being fully leveraged, limiting the advancement of MIMO signal detection research.

Our approach transforms the MIMO signal detection from a discrete problem to a continuous one, utilizing the efficient Hamiltonian Monte Carlo (HMC) method~\cite{DUANE1987216}.
We adopt a mixture of $t$-distributions as the prior distribution~\cite{Globecom} for this continuous framework.
Notably, \cite{Langevin} also employs a continuous prior distribution similar to that in our proposed method; however, they adopt the Langevin algorithm for the posterior distribution approximation.
The Langevin algorithm can be considered a special case of HMC where the Hamiltonian equations are solved numerically for only one step~\cite{LangevinandHamilton}.
Consequently, while \cite{Langevin} prioritizes computational efficiency, our proposed method focuses on the detection accuracy.

Furthermore, we consider a detection method for our approach in cases when channel coding, which is commonly employed in practical wireless communications, is applied.
To improve the coded performance, we integrate signal detection with error-correction decoding synergistically.
Additionally, the likelihood's temperature parameter is considered a random variable, enabling automatic optimization.
This treatment extends the likelihood from a normal density to a horseshoe density~\cite{2010HorseshoeBiometrrika}, as the temperature parameter follows a Cauchy distribution.

Theoretical analysis and simulations demonstrate our method's near-optimal detection performance with polynomial computational complexity.
Our innovative MIMO detection technique advances both the practical and theoretical domains of next-generation wireless networks.

This study builds upon the work of~\cite{Globecom}.
The primary enhancement is the application of channel coding, which entailed implementing substantial adaptations to the original framework.
Additionally, we conduct a detailed study to explain why the proposed method achieves better performance with the modulation order.

This paper is organized as follows:
Section~\ref{sec:Problem formulation} introduces the problem's stochastic formulation.
Section~\ref{sec:Previous work} provides an overview of established stochastic methods.
Sections~\ref{sec:Stochastic signal detection with a mixture of $t$-distributions prior} and \ref{sec:coded signal detection} detail the proposed approach for the uncoded and coded scenarios, respectively.
Section~\ref{sec:Numerical results and discussion} presents the outcomes of the theoretical analysis and numerical simulations.
Section~\ref{sec:Conclusions} concludes the paper with a summary of the key points.

The key notations utilized in this paper are presented below.
$\mathbb{C}$ represents the complex number field.
$\mathrm{Re}(\xbt)$ and $\mathrm{Im}(\xbt)$ indicate $\xbt$'s real and imaginary components, respectively.
$\hat{\xbt}$ represents the estimated value of $\xbt$.
$\text{E}[\xbt]$ denotes the expectation of $\xbt$.
$\| \xbt \|$ represents the Euclidean norm of $\xbt$.
$\zero$ and $\Ibt$ represent the zero vectors and the identity matrix, respectively.
$\Xbt^\top$ indicates the transpose of matrix $\Xbt$.
$\lfloor x \rfloor$ signifies the floor function, the largest integer less than or equal to $x$.
$\sim$ indicates that the left-side random variable follows the right-side probability distribution.
The proportionality between quantities is represented by $\propto$.
\color{black}
We assume a square quadrature-amplitude modulation (QAM) constellation, in which 4QAM is equivalent to quadrature phase-shift keying (QPSK).
The constellation set in real-valued representation is denoted by $\mathcal{S} = \{s_1, \ldots, s_k, \ldots, s_K\}$ (hence the modulation order is $K^2$).
\color{black}

\section{Problem formulation} \label{sec:Problem formulation}
\subsection{System Model}
We examine a full-stream MIMO configuration featuring $N$ transmit and $M$ receive antennas.
This MIMO system is described by the following relationship:
\begin{equation}
\qquad\qquad\qquad\qquad \ybt = \Hbt \ubt + \wbt, \quad \wbt \sim \mathcal{CN}(\zero, \sigma_w^2 \Ibt),
\label{eq:system}
\end{equation}
where $\ybt = [y_1, \ldots, y_M]^\top \in \mathbb{C}^M$ denotes the received symbol vector, $\Hbt \in \mathbb{C}^{M \times N}$ is the channel matrix, $\ubt = [u_1, \allowbreak \ldots, u_N]^\top \in \mathbb{C}^N$ represents the transmitted symbol vector, $\wbt = [w_1,  \ldots, w_M]^\top \in \mathbb{C}^M$ denotes the noise vector with variance $\sigma_w^2$, and $\mathcal{CN}$ denotes a circularly symmetric complex normal distribution.
For computation convenience, we decompose the complex numbers into real and imaginary parts as follows:
\rule[-1.7ex]{0ex}{0ex}%
$\ybt \rightarrow \left[ \begin{smallmatrix} \mathrm{Re}(\ybt) \\ \mathrm{Im}(\ybt) \end{smallmatrix}\right]$,
$\Hbt \rightarrow \left[ \begin{smallmatrix} \mathrm{Re}(\Hbt) & -\mathrm{Im}(\Hbt)\\ \mathrm{Im}(\Hbt) & \phantom{-}\mathrm{Re}(\Hbt) \end{smallmatrix}\right]$,
$\ubt \rightarrow \left[ \begin{smallmatrix} \mathrm{Re}(\ubt) \\ \mathrm{Im}(\ubt) \end{smallmatrix} \right]$, and
$\wbt \rightarrow \left[ \begin{smallmatrix} \mathrm{Re}(\wbt) \\ \mathrm{Im}(\wbt) \end{smallmatrix} \right]$.
\rule[-2ex]{0ex}{0ex}%
Note that the vector and matrix characters remain the same when $N$ and $M$ are doubled to $2N$ and $2M$, respectively.
In this study, we presume that the channel matrix $\Hbt$ and noise variance $\sigma_w^2$ are known.
Additionally, we assume that scramblers ensure the uniformly random transmission of symbols across the antennas, with known average power $P_{t}$.
The goal of signal detection is to estimate the transmitted symbol vector $\ubt$ from the received symbol vector $\ybt$.

From the stochastic perspective, Bayes' theorem (posterior distribution $\propto$ likelihood $\times$ prior distribution) yields
\begin{equation}
p(\ubt \mid \ybt) \propto p(\ybt \mid \ubt) p(\ubt).
\label{eq:Bayse}
\end{equation}
Therefore, stochastic signal detection aims to derive a point estimate from the posterior distribution $p(\ubt \mid \ybt)$.

\subsection{Likelihood}
Equation~\eqref{eq:system} demonstrates the following relationship:
\begin{equation}
p(\ybt \mid \ubt) = \mathcal{N}(\ybt; \Hbt \ubt, \sigma_w^2 \Ibt),
\label{eq:Likelihood}
\end{equation}
where $\mathcal{N}$ denotes the probability density of a real-valued normal distribution.
$\sigma_w$ represents the average noise amplitude.
In physics, this term is also referred to as the temperature parameter~\cite{Hassibi}.
When using MCMC to approximate posterior distributions, this value is sometimes intentionally altered from the original.
This is because it strongly influences the estimation.
However, it is generally difficult to optimize this parameter~\cite{Hassibi}.

\subsection{Prior Distribution}
The prior distribution, as shown in \eqref{eq:Bayse}, can be interpreted as a regularization term that calibrates the likelihood.
We can combine multiple prior distributions for different purposes.
For example, when applying two independent prior distributions simultaneously, we define them separately as $p_1(\ubt)$ and $p_2(\ubt)$ and subsequently consider their product, $p(\ubt)=p_1(\ubt)p_2(\ubt)$.
Similarly, this study considers multiple types of prior distributions, which will be explained in detail.

\begin{figure}[!tb]
\centering
\includegraphics[clip,width=0.8\linewidth]{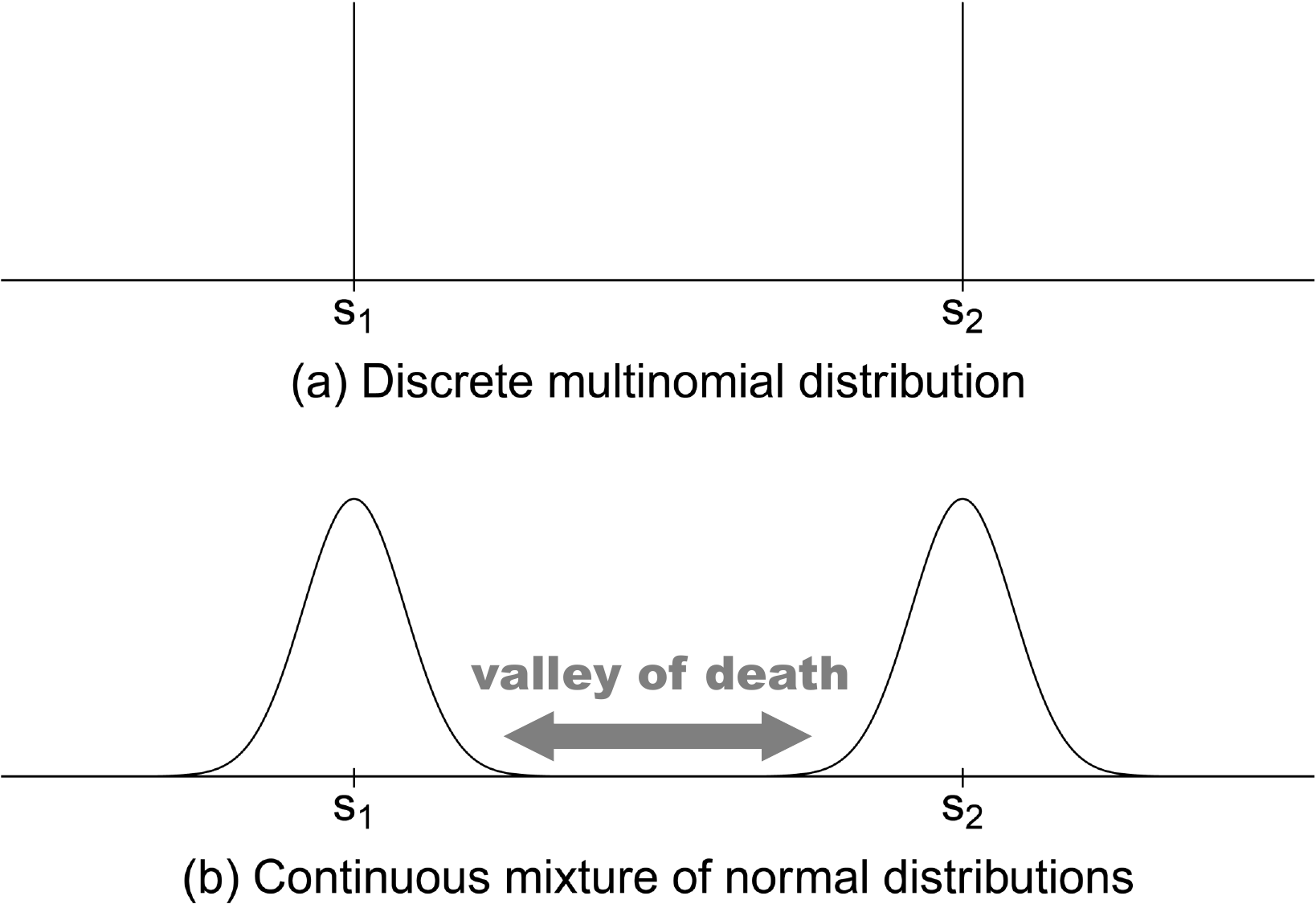}
\caption{Two examples of priors for binary phase-shift keying (BPSK).}
\label{fig:prior}
\end{figure}

The primary and essential prior distributions in signal detection represent the signal position, denoted as $p_1(\ubt)$.
For MIMO signal detection, a discrete multinomial prior distribution is commonly employed to represent potential discrete signal points.
This approach enhances the posterior estimation accuracy by prioritizing the transmission signal points, as shown in \eqref{eq:multinomial} (see Fig.~\ref{fig:prior} (a)):
\begin{equation}
p_1(\ubt) =  \prod^{2N}_{n=1} \frac{1}{K}\{\delta(u_n - s_1) + \cdots +\delta(u_n - s_K) \},
\label{eq:multinomial}
\end{equation}
where $\delta(x)$ is the unit probability mass at $x$.

Adopting a continuous prior distribution converts the discrete problem into a continuous problem.
Equation~\eqref{eq:mixturenormal} illustrates this approach using a mixture of normal distributions (see Fig.~\ref{fig:prior} (b)):
\begin{equation}
p_1(\ubt) = \prod^{2N}_{n=1}  \frac{1}{K}\{\mathcal{N}(u_n; s_1, \sigma) + \cdots + \mathcal{N}(u_n; s_K, \sigma) \},
\label{eq:mixturenormal}
\end{equation}
where $\sigma^2$ represents the normal distribution variance of each component, a crucial tuning parameter.
In our prior work~\cite{WPMC,MatsumuraRCS2020a,MatsumuraRCS2020b,AsumiRCS2021,AsumiRCS,AsumiRCS2023,KasugaGC2022}, we employed this type of prior distribution, optimizing $\sigma$ to minimize the bit error rate (BER) via preliminary searches.
The optimal $\sigma$ value represents an equilibrium point that maximizes search efficiency.
An overly high $\sigma$ value reduces search efficiency as irrelevant areas beyond the transmission signal points are explored.
Conversely, a very low $\sigma$ value limits component distribution overlap, hindering exploration across different signal points (the ``valley of death'' in Fig.~\ref{fig:prior} (b)).

Next, we may consider a secondary prior distribution to mitigate the noise enhancement effect~\cite{chockalingam2014large} in MIMO signal detection.
This prior distribution, denoted as $p_2(\ubt)$, is  a wide normal distribution centered at zero, whose effect is often referred to as ridge regularization~\cite{ridge}:
\begin{equation}
p_2(\ubt) = \mathcal{N}(\ubt; \zero, \sigma_\text{ridge}^2 \Ibt),
\label{eq:ridge}
\end{equation}
where $\sigma_\text{ridge}$ is a tuning parameter.
If we set $p(\ubt) = p_2(\ubt)$ alone, the posterior distribution can be derived analytically.
In this case, the mean of the posterior distribution becomes the well-known MMSE solution in the average setting $\sigma_\text{ridge}^2 = P_{t}$:
\begin{align}
p(\ubt \mid \ybt) & \propto p(\ybt \mid \ubt) p(\ubt) \notag \\
& = \mathcal{N}(\ybt; \Hbt \ubt, \sigma_w^2 \Ibt) \mathcal{N}(\ubt; \zero, P_{t} \Ibt) = \mathcal{N}(\ubt; \mubt, \Sigmabt),
\label{eq:MMSE}
\end{align}
where $\mubt = \Sigmabt \Hbt^\top \ybt$ and $\Sigmabt = (\Hbt^\top \Hbt + \sigma_w^2/P_{t} \Ibt)^{-1}$.

\subsection{Posterior Distribution}
\begin{figure}[!tb]
\centering
\includegraphics[draft=false,clip,width=0.9\linewidth]{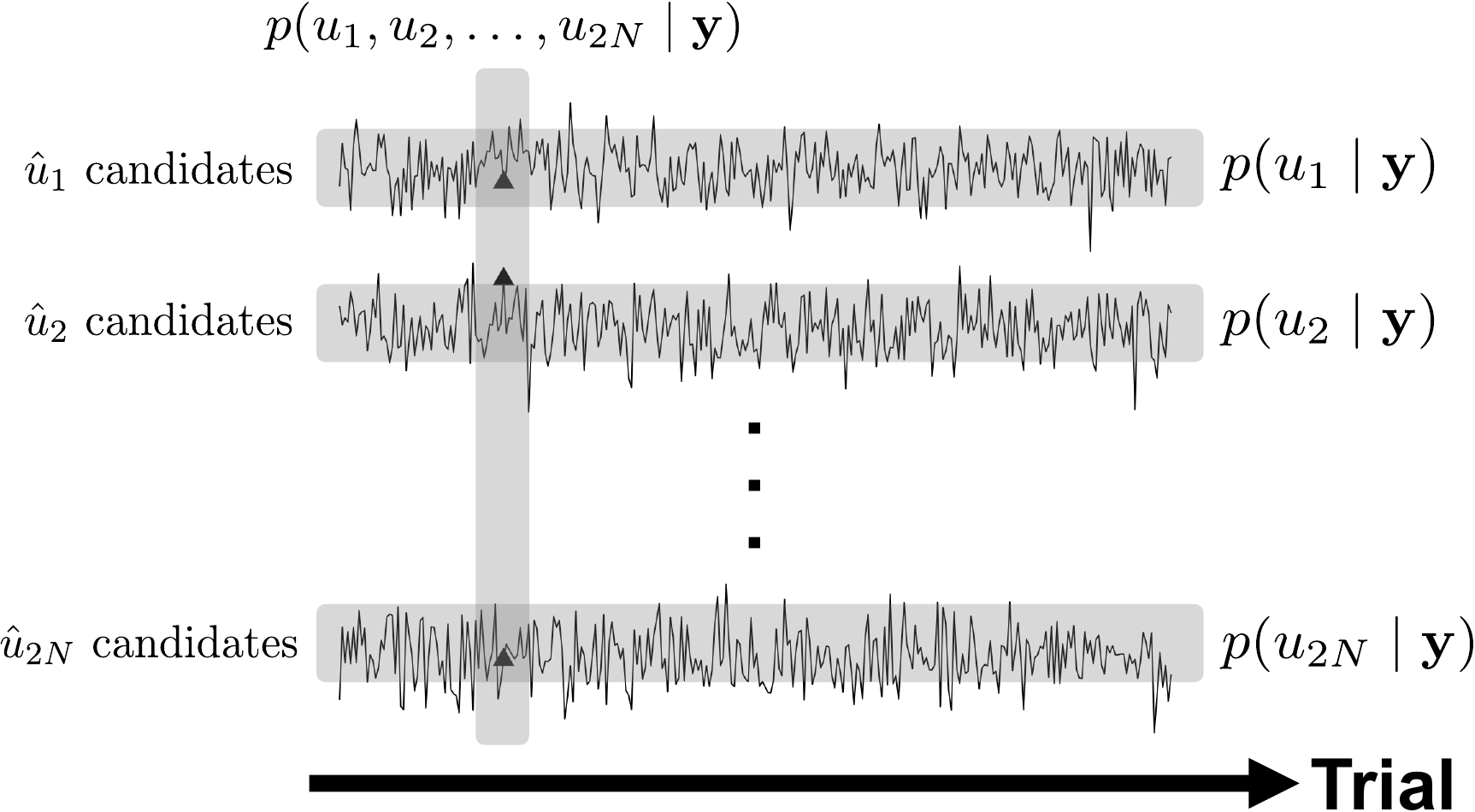}
\caption{Criteria for selecting a single one from soft-value symbol candidates.}
\label{fig:Posterior}
\end{figure}
Assuming a mixture distribution as the prior distribution prevents closed-form analytical solutions for the posterior distribution~\cite{bishop:2006:PRML}.
Deriving the posterior distribution requires numerical approximation algorithms.
The point estimate of the posterior distribution, $\hat{\ubt}$, may yield various candidates based on the chosen approximation algorithm.
Moreover, $\hat{\ubt}$ might deviate from the actual transmission signal point.
A continuous prior distribution may lead to explorations outside the discrete signal points, inducing potential deviations in the estimates.

To address these issues,
from the candidates for the soft-value symbol $\hat{\ubt}$, we select a single one according to a certain criterion and quantize it to the nearest signal point if a hard decision is needed.
We can consider several criteria for selecting the final estimate.
The primary approach lies in the joint posterior distribution $p(u_1, u_2, \ldots, u_{2N} \mid \ybt)$ (see Fig.~\ref{fig:Posterior}).
First, we calculate the likelihood $p(\ybt \mid \tilde{\ubt})$ for each quantized $\tilde{\ubt}$ from $\hat{\ubt}$. Thereafter, we select the $\hat{\ubt}$ corresponding to the $\tilde{\ubt}$ which yields the highest likelihood.
We refer to this as the joint posterior-based point estimator.
The secondary approach lies in the marginal posterior distribution $p(u_n \mid \ybt)$ (see Fig.~\ref{fig:Posterior}).
We select the set \{$\text{E}[{u_1} \mid \ybt], \text{E}[{u_2} \mid \ybt], \ldots, \text{E}[u_{2N} \mid \ybt]$\} as the final estimate.
We refer to this as the marginal posterior-based point estimator.
Generally, while the joint distribution contains detailed information, it is averaged out during marginalization.
Therefore, although estimation should ideally be based on the joint distribution, the well-behaved marginal distribution can also be beneficial under noisy conditions.
We primarily employ the joint posterior-based point estimator, while also employing the marginal posterior-based point estimator, depending on the situation.

\section{Previous work} \label{sec:Previous work}
\subsection{MGS Method~\cite{MGS}} \label{subsec:MGS}
The MGS method estimates the posterior distribution through extensive sample generation utilizing Gibbs sampling.
The MGS method enhances efficiency by refreshing the initial search values with probability $1/(2N)$, thereby avoiding local optima traps.
With this approach, the Markov chains are virtually parallelized.
Additionally, \cite{MGS} proposes a multiple-restart (MR) technique, which runs several MGS processes with varied initial Markov chain values and thereafter selects the highest likelihood result.
\color{red}
\marginpar{{\footnotesize [R1.5]\par[R3.4]}}%
Let $L_\text{MGS}$ represent the total number of Markov chain steps and $\text{MR}_\text{MGS}$ denote the number of MR iterations.
According to~\cite{MGS}, sufficient performance can be achieved with $L_\text{MGS}=8N K$ and $\text{MR}_\text{MGS}=50$.
Therefore, we set $L_\text{MGS}=8N K$ in this study, while $\text{MR}_\text{MGS}$ is determined later for fair comparison with other methods.
\color{black}
The MGS method's complexity per Markov chain step is $\mathcal{O}(2M2N)$, primarily due to the likelihood computation.
The overall complexity becomes $\text{MR}_\text{MGS} \times \mathcal{O}(L_\text{MGS} 2M2N)$.

\subsection{EP Method~\cite{EP}}
The EP method estimates the posterior distribution with an uncorrelated multivariate normal distribution $q(\ubt)$.
The method uses the EP algorithm to find the parameters of $q(\ubt)$.
The approach iteratively refines the mean and variance parameters, using the final mean to estimate the transmitted symbols.
\cite{EP} suggests setting the total number of iterations $L_\text{EP}$ to at most 10 to achieve optimal detection performance.
Therefore, we set $L_\text{EP}=10$ in this study.
The per-iteration complexity of the EP method is $\mathcal{O}((2N)^3)$, primarily due to its internal matrix inversion operation.
The final complexity reaches $\mathcal{O}(L_\text{EP} (2N)^3)$.

\color{red}
\subsection{MH Gradient Descent (MHGD) Method~\cite{GradientBasedMH}}
\marginpar{{\footnotesize [R2.2]}}%
The MHGD method estimates the posterior distribution based on the MH algorithm, which is a type of MCMC.
The method leverages likelihood gradient information to enhance search efficiency, with quantization to the nearest constellation point at each Markov chain step.
Referring to~\cite{GradientBasedMH}, we set the number of Markov chain steps as $L_\text{MHGD}=8 \times 2N$ to achieve sufficient performance in this study.
The computational complexity is $\text{MR}_\text{MHGD} \times \mathcal{O}(L_\text{MHGD} (2N)^2)$, where $\text{MR}_\text{MHGD}$ denotes the number of MR iterations.

\subsection{Annealed Langevin Method~\cite{Langevin}}
\marginpar{{\footnotesize [R2.2]}}%
The Langevin method solves the Hamiltonian equations for only a single step and always accepts the proposal, and is therefore regarded as an optimization technique~\cite{LangevinandHamilton}.
In~\cite{Langevin}, a mixture of normal distributions is assumed as the prior, and its scale parameter is gradually decreased during the iterative process of obtaining the posterior distribution.
Based on~\cite{Langevin}, we adopt $T_\text{A} = 20$ annealing steps and $L_\text{Lang}=70$ inner optimization iterations to achieve satisfactory performance.
The computational complexity is $\text{MR}_\text{Lang} \times \mathcal{O}(T_\text{A} L_\text{Lang}(2N)^2 + 2N2M \times \min(2N, 2M))$, accounting for the use of singular value decomposition (SVD), where $\text{MR}_\text{Lang}$ denotes the number of MR iterations.

\subsection{OAMP-Net2~\cite{OAMPNet2}}
\marginpar{{\footnotesize [R3.8]}}%
The underlying OAMP algorithm~\cite{OAMP} first approximates the linear estimation residual with a unimodal normal distribution, followed by nonlinear operations incorporating discrete symbol constellations to refine the regularization term in the linear estimation.
By iterating this process, the estimation accuracy is improved.
The hyperparameters in OAMP-Net2 depend on factors such as $\Hbt$ and are pre-trained to further improve the accuracy.
Following~\cite{OAMPNet2}, we set the number of training layers as $T_\text{O} = 10$ to achieve sufficient performance in this study.
The computational complexity is $O(T_\text{O} (2N)^3)$ when the hyperparameters are pre-trained.
\color{black}

\section{Proposed signal detection for the uncoded case} \label{sec:Stochastic signal detection with a mixture of $t$-distributions prior}
\subsection{System Diagram}
\begin{figure}[!tb]
\raggedright
\includegraphics[draft=false,clip,width=0.9\linewidth]{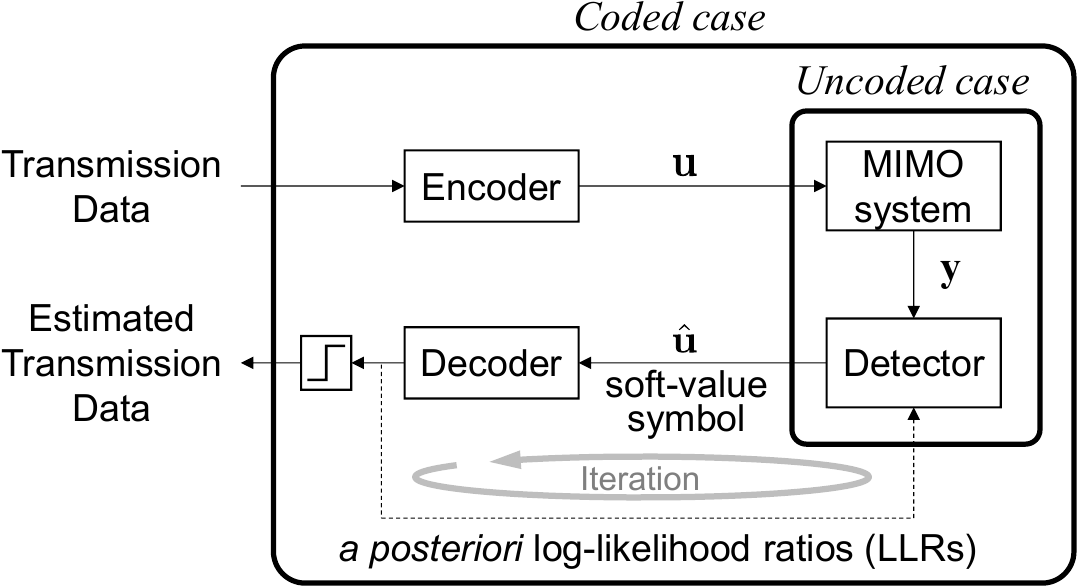}
\caption{System diagram for the uncoded and coded cases.}
\label{fig:Systemmodel_uncoded}
\end{figure}

Fig.~\ref{fig:Systemmodel_uncoded} shows the system diagram of the proposed detector.
The inner part represents the uncoded case, where the input is $\ubt$ and the output is a soft-value symbol $\hat{\ubt}$ (final estimate).
For hard decisions, the output is quantized to the nearest constellation point.

\subsection{Likelihood}
For the uncoded case, the likelihood is given by \eqref{eq:Likelihood}.

\subsection{Prior Distribution}
For the uncoded case, we assume that $p(\ubt) = p_1(\ubt)$ by considering only the primary $p_1(\ubt)$.

We adopt a mixture of $t$-distributions as the prior distribution for signal position, expressed in \eqref{eq:mixturet}:
\begin{equation}
p_1(\ubt) = \prod^{2N}_{n=1} \frac{1}{K} \{\mathcal{T}(u_n; s_1, \sigma, \nu) + \cdots + \mathcal{T}(u_n; s_K, \sigma, \nu) \},
\label{eq:mixturet}
\end{equation}
where $\mathcal{T}$ represents a real-valued $t$-distribution density~\cite{bishop:2006:PRML} with tunable parameters $\sigma$ (scale) and $\nu$ (degrees of freedom).
We utilize the optimal values for these parameters, determined through preliminary searches.
The $t$-distribution features a sharper peak and heavier tails than the normal distribution with identical location and scale parameters.
Consequently, $t$-distributions enable a more exhaustive exploration around the transmission signal points than normal distributions.
Furthermore, they facilitate the exploration of other potential signal points by traversing the inter-point valley of death.
This approach offers superior exploration compared to our previous normal distribution mixture~\cite{WPMC,MatsumuraRCS2020a,MatsumuraRCS2020b,AsumiRCS2021,AsumiRCS,AsumiRCS2023,KasugaGC2022}.

\subsection{Posterior Distribution}
In the uncoded case, we consider only the primary joint posterior-based point estimator.

Next, we focus on algorithms that numerically approximate the posterior distribution.
Signal detection using a continuous prior distribution leads to a continuous posterior distribution, attributed to the continuous likelihood.
This allows the use of powerful continuous-domain algorithms for the posterior distribution approximation.
Previously, we evaluated Newton's method, automatic differentiation variational inference (ADVI~\cite{ADVI}), and HMC as approximation algorithms using normal mixture priors~\cite{AsumiRCS}.
The results indicated HMC's superior signal-detection performance at comparable computational complexities.
HMC's auxiliary momentum variables, randomly redrawn at each Markov chain step, appear to mitigate the local optima issues often encountered in Newton's method and ADVI.
Therefore, we utilize the HMC method to approximate the posterior distribution in our research.

HMC, an MCMC variant, employs Markov chain principles similar to the MGS and MHGD methods.
Therefore, HMC samples regions with high posterior probability densities to approximate the posterior distribution.
The sampling efficiency of HMC typically surpasses that of other MCMC methods, attributed to its innovative application of Hamiltonian mechanics.
The key aspects of the HMC method are outlined below.
First, the HMC introduces an auxiliary momentum variable $\rbt = \mathrm{d}\ubt/\mathrm{d}\tau$, where $\tau$ denotes virtual time, alongside the estimated variable $\ubt$.
Further, the method considers the potential energy $U(\ubt) = -\ln(p(\ubt \mid \ybt))$ and kinetic energy $K(\rbt) = 1/2 \|\rbt\|^2$.
The Hamiltonian representing the total system energy is defined as
\begin{equation}
H(\ubt, \rbt) = U(\ubt) + K(\rbt).
\label{eq:Hamiltonian}
\end{equation}
Hamilton's equations are now formulated as two ordinary differential equations:
\begin{equation}
\begin{split}
\frac{\mathrm{d}\ubt}{\mathrm{d}\tau} & =  \frac{\partial H(\ubt, \rbt)}{\partial\rbt} = \rbt, \\
\frac{\mathrm{d}\rbt}{\mathrm{d}\tau} & = -\frac{\partial H(\ubt, \rbt)}{\partial\ubt} = -\frac{\partial U(\ubt)}{\partial\ubt}.
\label{eq:hamiltonsde}
\end{split}
\end{equation}
Algorithm~\ref{alg:hmc} illustrates the HMC sampling process.
\begin{algorithm}[!tb]
\caption{HMC sampling}
\label{alg:hmc}
{\footnotesize
\begin{algorithmic}[1]
\State Initialize $\ubt$ at random
\For{$l = 1, \dots, L_\text{HMC}$}
	\State Draw $\rbt$ from $\mathcal{N}(\zero, \Ibt)$
	\State \parbox[t]{1.0\linewidth}{Numerically solve Hamilton's equations~\eqref{eq:hamiltonsde} to obtain $\ubt^\prime$ and $\rbt^\prime$}
	\State \parbox[t]{1.0\linewidth}{Update $\ubt \leftarrow \ubt^\prime$ with probability min$[1,  \exp\{H(\ubt, \rbt) - H(\ubt^\prime, \rbt^\prime)\}]$}
	\State \parbox[t]{0.9\linewidth}{Regard the updated $\ubt$ as a sample from the posterior distribution $p(\ubt \mid \ybt)$}
\EndFor
\end{algorithmic}
}
\end{algorithm}
The Hamiltonian remains constant according to \eqref{eq:Hamiltonian}.
Consequently, the substantial changes in the momentum $\rbt$ significantly impact the sample value $\ubt$.
Meanwhile, Algorithm~\ref{alg:hmc} shows that most proposals ($\ubt^\prime$) are accepted with a probability of almost one, except for numerical error cases.
These features contribute to HMC's superior sampling efficiency compared to those of other MCMC methods.

\color{red}
\marginpar{{\footnotesize [R1.1]\par[R3.1]}}%
To numerically solve \eqref{eq:hamiltonsde} within a single step of the Markov chain, we need to perform $L$ internal iterations where $\tau$ is advanced by small increments $\varepsilon$ and the values are evaluated at each virtual time.
In this process, the leapfrog method is commonly employed rather than Euler's method to minimize cumulative numerical errors~\cite{leapflog}.
The appropriate values of $\varepsilon$ and $L$ vary depending on the problem.
For instance, if $\varepsilon$ is too small, similar proposals occur redundantly. If $L$ is too large, the trajectory makes a U-turn to previously evaluated regions, resulting in wasted computations.
The no-U-turn sampler (NUTS)~\cite{NUTS} was proposed as a milestone automated method to address these issues, which we also adopt in this study.
The overview of NUTS is as follows.
First, $\varepsilon$ is adaptively adjusted during the warm-up phase using dual averaging~\cite{Nesterov} to approach the target acceptance rate of proposals.
For $L$, the method employs a doubling strategy borrowed from slice sampling~\cite{slicesampling}, in which the trajectory is progressively extended using a binary-tree structure. The algorithm performs tentative sampling until the U-turn condition is detected, then randomly selects one point from the constructed tree according to the target distribution.
\color{black}

The computational complexity of the HMC method is primarily determined by the term $(\Hbt^\top \Hbt)\ubt$.
This term emerges from the log-likelihood component in the differentiation of the log-posterior probability density. Since $\Hbt$ is known, $(\Hbt^\top \Hbt)$ only needs to be computed once.
Consequently, the HMC's per-step computational complexity is $\mathcal{O}(L (2N)^2)$.
While $L$ may vary with the problem, we typically set $L=10$ in this study, following an example in~\cite{BDA3}.
The total complexity, given $L_\text{HMC}$ Markov chain steps, becomes $\mathcal{O}(10 L_\text{HMC} (2N)^2)$.

\color{red}
\subsection{Summary of the Proposed Method}
\marginpar{{\footnotesize [R2.8]\par[R3.7]}}%
Table~\ref{tbl:uncoded_summary} summarizes the proposed method.
\begin{table}[!tb]
\centering
\caption{Proposed method summary for uncoded case}
\label{tbl:uncoded_summary}
\footnotesize
\setlength{\tabcolsep}{4pt}
\begin{tabular}{ll|c}
\hline
\multicolumn{2}{c|}{Diagram} & Fig.~\ref{fig:Systemmodel_uncoded} (inner part) \\
\hline
\multicolumn{2}{c|}{Likelihood} & Normal:~\eqref{eq:Likelihood}\\
\hline
Prior & $p(\mathbf{u})$ & $p_1(\mathbf{u})$:~\eqref{eq:mixturet}\\
\hline
\multirow{2}{*}{Posterior} 
& Stats & \begin{tabular}[c]{@{}c@{}}Joint posterior-based point estimator\\$p(\mathbf{u}|\mathbf{y})$\end{tabular} \\
\cline{2-3}
& Algo & HMC \\
\hline
\multicolumn{2}{c|}{Complexity} & $\mathcal{O}(10L_{\text{HMC}}(2N)^2)$ \\
\hline
\end{tabular}
\end{table}
According to \eqref{eq:Bayse}, the relationship among the likelihood, prior, and posterior is as follows:
$
p(\ubt \mid \ybt) \propto p(\ybt \mid \ubt) \times p(\ubt)
= \text{normal likelihood} \times \text{prior of $\ubt$}.
$
\color{black}

\section{Proposed signal detection for the coded case} \label{sec:coded signal detection}
\subsection{System Diagram}
Fig.~\ref{fig:Systemmodel_uncoded} shows the system diagram of the proposed detector.
The inner uncoded case is expanded to the outer coded case by incorporating an encoder and a decoder.
\marginpar{{\footnotesize [R2.4]\par[R3.5]}}%
\color{red}
Error correction is achieved using low-density parity-check (LDPC) codes, which have a proven track record in 5G~\cite{5GNRLDPC}.
\color{black}
The decoder's input is the soft-value symbol $\hat{\ubt}$ (final estimate) from the detector, based on the posterior distribution.
The decoder's output is bit-wise \textit{a posteriori} log-likelihood ratios (LLRs), which are the final values after the LDPC decoding internal iterations.
Feeding these LLRs back to the detector enables iterative detection and decoding, which enhances estimation performance~\cite{SOAV1}.
Accordingly, we employ the LLR feedback approach in this study.
While the initial detection and decoding do not involve LLR feedback, joint detection and decoding with the LLR feedback are implemented in subsequent phases.

\subsection{Likelihood}\label{subsec:codedLik}
In the coded case, we consider an extension of \eqref{eq:Likelihood} for the joint detection and decoding with LLR feedback.
This extension aims to further enhance the performance of our proposed method utilizing the LLR.

As mentioned earlier, when using MCMC to approximate the posterior distribution, we can improve the detection performance by intentionally modifying the temperature parameter from the average noise amplitude.
However, this optimization task is generally challenging.
Therefore, this study proposes treating the temperature parameter as a random variable for automatic optimization.
Specifically, we extend the formulation by multiplying the average noise amplitude by a positive coefficient:
\begin{equation}
\sigma_w \rightarrow \lambda_n \sigma_w. \quad (n = 1, \ldots, 2N) 
\label{eq:temperature_parameter}
\end{equation}
Here, we consider $\lambda_n$ to be an independent random variable.
This approach maintains the average noise amplitude as a reference while easily accommodating significantly different values when necessary.
Henceforth, let $\lambdabt = [\lambda_1, \ldots, \lambda_{2N}]^\top$.

We can assume various probability distributions for $\lambda_n$.
The most basic one is a uniform distribution limited to positive support.
However, uniform distributions may lead to insufficient estimation performance when computational resources are finite. This is because all positive regions are explored with equal weight.
We propose using a Cauchy distribution as follows:
\begin{equation}
\lambda_n \sim \mathcal{C}^{+}(0, \sigma_\text{Cauchy}),
\label{eq:halfCauchy}
\end{equation}
where $\mathcal{C}^{+}$ denotes a Cauchy distribution limited to positive support, with the location parameter 0 and the scale parameter $\sigma_\text{Cauchy}$.
The optimal value for $\sigma_\text{Cauchy}$ is determined through preliminary searches.
Under this setup, the likelihood extends from $p(\ybt \mid \ubt)$ to $p(\ybt \mid \ubt, \lambdabt)$, which follows a normal density.
Consequently, the standard $p(\ybt \mid \ubt)$, after marginalizing $\lambdabt$ out, follows a horseshoe density~\cite{2010HorseshoeBiometrrika}.
The horseshoe distribution, characterized by a thin peak and heavy tails, is suitable for our problem.
It typically takes values around a specific low value but occasionally produces significantly deviant values.
This distribution enables targeted exploration around the peak (near average noise amplitude) and the tails (far from average noise amplitude).
Thus, it is expected to enhance the exploration efficiency when computational resources are finite, improving the estimation performance.
Originally, the horseshoe distribution was proposed as a prior distribution for the regression coefficients in sparse regression~\cite{2010HorseshoeBiometrrika}.
In this study, however, we apply it to the likelihood rather than the prior distribution for transmission symbols.

The introduction of the new random variable $\lambda_n$ increases the computational complexity compared with that in the uncoded case.
Processing $\lambdabt$ in the likelihood within the HMC framework incurs $\mathcal{O}(2N)$ additional operations per step, accounting for their mutual independence.
Nevertheless, this additional load is insignificant compared to the previously clarified per-step complexity of $\mathcal{O}(L (2N)^2)$ for the uncoded case.
Therefore, this increase does not substantially affect the total computational expenditure.

\subsection{Prior Distribution}
As described in Section~\ref{subsec:codedLik}, in the coded case, we may consider $\lambdabt$ in addition to $\ubt$.
In this case, the prior distribution is extended from $p(\ubt)$ to $p(\ubt, \lambdabt)$.
Since $\ubt$ and $\lambdabt$ are mutually independent, $p(\ubt, \lambdabt) = p(\ubt) \times p(\lambdabt)$.
For $p(\ubt)$, we consider $p_1(\ubt)$ and $p_2(\ubt)$, setting $p(\ubt)=p_1(\ubt)p_2(\ubt)$.
Here, $p_1(\ubt)$ is extended from \eqref{eq:mixturet} to accommodate the LLR feedback from the decoder.
$p_2(\ubt)$ is introduced to improve the performance at low signal-to-noise ratios (SNRs).
Furthermore, when considering $\lambdabt$, we include an additional prior distribution $p(\lambdabt)$ based on \eqref{eq:halfCauchy}.

First, regarding $p_1(\ubt)$, the term $1/K$ in \eqref{eq:mixturet} implies that an equal occurrence frequency is adopted for each symbol.
To consider joint detection and decoding with the LLR feedback, we extend the definition of $p_1(\ubt)$ to express the uneven symbol occurrence frequencies:
\begin{equation}
p_1(\ubt) = \prod^{2N}_{n=1} \{\omega_1 \mathcal{T}(u_n; s_1, \sigma, \nu) + \cdots + \omega_K \mathcal{T}(u_n; s_K, \sigma, \nu)\},
\label{eq:mixturet_enhance}
\end{equation}
where $\sum_{k=1}^K \omega_k = 1$.
Although $\omega_k$ is set to $1/K$ for the initial detection, it becomes an LLR-dependent value for the subsequent joint detection and decoding.
The relationship between $\omega_k$ and the LLR feedback from the decoder is described below.
Let the bit sequence representation of a symbol be $b_1 \cdots b_d \cdots b_D$.
Here, $D = \log_2 K$.
Given that the LLR for $b_d$ is denoted as $\Lambda_d$, where $\Lambda_d := \ln \frac{p(b_d = `1\text{'} \mid \ybt)}{p(b_d = `0\text{'} \mid \ybt)} = \ln \frac{p(\ybt \mid b_d = `1\text{'})}{p(\ybt \mid b_d = `0\text{'})}$, we obtain the following equations:
\begin{align}
p(b_d =`0\text{'} \mid \ybt) & = \phantom{e^{\Lambda_d}} 1 /(1+e^{\Lambda_d}), \label{eq:LLR0} \\
p(b_d =`1\text{'} \mid \ybt) & = \phantom{1} e^{\Lambda_d} /(1+e^{\Lambda_d}). \label{eq:LLR1}
\end{align}
Therefore, we can obtain the weight $\omega_k$ by multiplying \eqref{eq:LLR0} and \eqref{eq:LLR1} according to the bit sequence of the $k$th symbol.
For 16QAM, $D=\log_2 \sqrt{16} =2$, yielding the bit sequence $b_1 b_2$.
For a symbol with the bit sequence $b_1 b_2 = 01$, the weight is $p(b_1 = `0\text{'} \mid \ybt) p(b_2 = `1\text{'} \mid \ybt)$.

Next, we set the variance parameter in $p_2(\ubt)$ given by \eqref{eq:ridge} as follows:
\begin{equation}
\sigma_\text{ridge}^2 = \sigma_w^{2} / \left[\text{max}\{\text{SVD}(\Hbt)\}^2 / \lambda_\text{ridge}\right],
\label{eq:ridge_sigma}
\end{equation}
where $\text{SVD}(\Hbt)$ represents the singular values of $\Hbt$, and $\lambda_\text{ridge} >1$ is a constant.
The optimal value of $\lambda_\text{ridge}$ is determined through preliminary searches.
Unlike MMSE, wherein $\sigma_\text{ridge}^2$ is set to the average transmission power $P_{t}$, our method depends on $\sigma_w$ and $\Hbt$.
This improves the adaptability to the SNR and individual channels, enhancing the detection performance.

Finally, when considering $\lambdabt$, we include
\begin{equation}
p(\lambdabt)=\prod_{n=1}^{2N} p(\lambda_n) = \prod_{n=1}^{2N} \mathcal{C}^{+}(\lambda_n; 0, \sigma_\text{Cauchy})
\label{eq:halr_Cauchy_prior_setting}
\end{equation}
based on \eqref{eq:halfCauchy}.

The introduction of $p_2(\ubt)$ and $p(\lambdabt)$ increases the computational complexity compared to the complexity recorded in the uncoded case.
The most computationally intensive aspect in $p_2(\ubt)$ is deriving the singular values.
We need to perform this calculation only once initially because we assume that $\Hbt$ is known.
This calculation typically adds $\mathcal{O}(2M 2N \times \text{min}\{2M, 2N\})$ operations.
Regarding $p(\lambdabt)$, the additional per-step complexity within the HMC is $\mathcal{O}(2N)$, due to the mutual independence of $\lambda_n$.
However, this increase is negligible compared to the previously identified per-step computational complexity of $\mathcal{O}(10 (2N)^2)$ for the uncoded case.
Therefore, it does not dominate the overall computational cost.

\subsection{Posterior Distribution}
In the coded case, in addition to the joint posterior-based point estimator, we consider the marginal posterior-based point estimator.
Specifically, we apply the marginal posterior-based point estimator for the initial detection without the LLR feedback.
We employ the joint posterior-based point estimator for subsequent joint detection and decoding with the LLR feedback.
This approach aims to improve the detection performance at low SNRs in the initial phase without LLR assistance.
In subsequent phases, the approach aims to achieve the best estimation with the highest likelihood aided by the LLR.

To approximate the posterior distribution, we apply the effective HMC algorithm, as in the uncoded case.

Notably, when considering $\lambdabt$, the joint posterior extends from $p(\ubt \mid \ybt)$ to $p(\ubt, \lambdabt \mid \ybt)$.
Our main objective remains the estimation of $\ubt$.
$\lambdabt$ serves only as the auxiliary variable for improving the estimation accuracy.
Thus, we consider $p(\ubt \mid \ybt)$ as the joint posterior, marginalizing out $\lambdabt$.
This marginalization can be achieved simply by discarding the $\lambdabt$ samples obtained simultaneously, thereby maintaining the computational complexity.

\color{red}
\subsection{Summary of the Proposed Method}
\marginpar{{\footnotesize [R2.8]\par[R3.7]}}%
Table~\ref{tbl:coded_summary} summarizes the proposed method.
\begin{table}[!tb]
\centering
\caption{Proposed method summary for coded case}
\label{tbl:coded_summary}
\footnotesize
\setlength{\tabcolsep}{3pt}
\begin{tabular}{ll|c|c}
\multicolumn{2}{l}{} & \multicolumn{2}{c}{\hspace{1em} Initial phase \hspace{5em} Subsequent phases} \\
\hline
\multicolumn{2}{c|}{Diagram} & \multicolumn{2}{c}{Fig.~\ref{fig:Systemmodel_uncoded}} \\
\hline
\multicolumn{2}{c|}{Likelihood} & Normal: \eqref{eq:Likelihood} & Horseshoe: \eqref{eq:Likelihood}, \eqref{eq:temperature_parameter}, \eqref{eq:halfCauchy}\\
\hline
\multirow{3}{*}{\begin{tabular}[c]{@{}l@{}}Prior \end{tabular}} 
& \multirow{2}{*}{$p(\mathbf{u})$} & \multicolumn{2}{c}{$p_1(\mathbf{u})p_2(\mathbf{u})$ \quad ($p_1(\mathbf{u})$: \eqref{eq:mixturet_enhance}, $p_2(\mathbf{u})$: 
\eqref{eq:ridge})} \\
\cline{3-4}
& & $\omega_k$: uniform & $\omega_k$: LLR-dependent \\
\cline{2-4}
& $p(\lambdabt)$ & --- &~\eqref{eq:halr_Cauchy_prior_setting}\\
\hline
\multirow{2}{*}{\begin{tabular}[c]{@{}l@{}}Posterior \end{tabular}} 
& Stats & \begin{tabular}[c]{@{}c@{}}Marginal posterior-based\\point estimator\\$\{\text{E}[{u}_1|\mathbf{y}], \ldots, \text{E}[{u}_{2N}|\mathbf{y}] \} $\end{tabular} 
& \begin{tabular}[c]{@{}c@{}}Joint posterior-based\\point estimator\\$p(\mathbf{u}|\mathbf{y}) = \int p(\mathbf{u}, \lambdabt|\mathbf{y})\mathrm{d}\lambdabt$\end{tabular} \\
\cline{2-4}
& Algo & \multicolumn{2}{c}{HMC} \\
\hline
\multicolumn{2}{c|}{Complexity} & \multicolumn{2}{c}{$\mathcal{O}(10L_{\text{HMC}}(2N)^2 + 2M2N \times \min\{2M, 2N\})$} \\
\hline
\end{tabular}
\end{table}
The relationship among the likelihood, prior, and posterior is given by \eqref{eq:Bayse} when not considering $\lambdabt$.
When considering $\lambdabt$, in principle we need to augment $\ubt$ to \{$\ubt$, $\lambdabt$\}, but ultimately we integrate out $\lambdabt$:
$
p(\ubt \mid \ybt) = \int p(\ubt, \lambdabt \mid \ybt) \, \mathrm{d}\lambdabt
\propto \int p(\ybt \mid \ubt, \lambdabt) \, p(\lambdabt \mid \ubt) \, \mathrm{d} \lambdabt \times p(\ubt)
= p(\ybt \mid \ubt) \times p(\ubt)
= \text{horseshoe likelihood} \times \text{prior of $\ubt$}.
$
\color{black}

\section{Numerical results and discussion} \label{sec:Numerical results and discussion}
Our proposed technique achieves near-optimal MIMO signal detection while maintaining polynomial computational complexity.
We validate these claims through theoretical complexity analyses and extensive signal-detection simulations.
\color{red}
\marginpar{{\footnotesize [R1.1]\par[R3.1]}}%
The source code used for the simulations is available at {\url{https://github.com/hagijyun/MIMO_HMC_detector}}.
\color{black}

\subsection{Settings in Numerical Analysis}
\subsubsection{Common Assumptions}
The study considers the typical conditions for massive MIMO systems, as shown in Table~\ref{tbl:simulationassumption}.
\begin{table}[!tb]
\caption{Common assumptions in numerical analyses}
\label{tbl:simulationassumption}
\centering
\begin{tabular}{lp{4.5cm}} \hline
\noalign{\vspace{0.3mm}} 
Item & Setting \\ \hline
\noalign{\vspace{0.3mm}} 
Trials & 5000 transmission symbol vectors (uncoded) and 100 codewords (coded)\\
Number of antennas & $N = M = 96$ \\
Modulation order & QPSK, 16QAM, and 64QAM \\
Symbol mapping & Gray code \\
Average transmission power & $P_{t} = 1/2 $ \\
Fading & Quasi-static Rayleigh \\
Channel correlation & Kronecker model (correlation coefficient $\rho = 0,\,0.5$) \\
Channel coding& Uncoded and coded cases\\ \hline
\end{tabular}
\end{table}
\color{red}
\marginpar{{\footnotesize [R2.4]\par[R3.5]}}%
Furthermore, we configured the LDPC encoding and decoding settings, as shown in Table \ref{tbl:turbo}.
\color{black}
These settings ensure that the error-correction code delivers adequate performance.
\begin{table}[!tb]
\caption{Channel coding and decoding settings}
\label{tbl:turbo}
\centering
\begin{tabular}{lp{5.8cm}} \hline
\noalign{\vspace{0.3mm}}
Item                               & Setting \\ \hline
\noalign{\vspace{0.3mm}}
LDPC coding                   & 3GPP 5G NR base graph 1~\cite{5GNRLDPC} (lifting size $Z$: 96) \\
LDPC decoding              & algorithm: sum-product (max iterations: 5) \\
Code rate                       & 0.3235 \\ 
Code length [bits]	       & 6528 \\ \hline
\end{tabular}
\end{table}
BER plots are excluded for simulations yielding error-free results.

\subsubsection{Parameters of Proposed Method}
Parameters $\sigma$ and $\nu$ for $t$-distribution, $\sigma_\text{Cauchy}$ for Cauchy distribution, and $\lambda_\text{ridge}$ for ridge regularization were optimized via a preliminary search (see Table~\ref{tbl:N_and_t_parameters}).
\begin{table}[!tb]
\caption{Parameters determined through the preliminary search}
\label{tbl:N_and_t_parameters}
\centering
\begin{tabular}{ccccc} \hline
\noalign{\vspace{0.3mm}}
		& \multicolumn{2}{c}{Mixture of $t$-distributions}													\\ \cline{2-3}
		&	$\sigma$		 		& $\nu$ 				& $\sigma_\text{Cauchy}$	& $\lambda_\text{ridge}$	\\ \hline
\noalign{\vspace{0.3mm}}
QPSK	&	0.1242				& 1.8				& 3.5					& \phantom{0}15			\\
16QAM	&	0.0621				& 1.8 				& 5.0					& \phantom{0}62			\\
64QAM	&	0.0531				& 2.5 				& 3.0					& 230					\\ \hline
\end{tabular}
\end{table}
Aligned with the virtual parallelization concept in the MGS, we configured each simulated Markov chain to comprise $2N$ steps.
The number of parallel Markov chains was set to $\lfloor 1000/(2N) \rfloor$, ensuring sufficient performance~\cite{Globecom}.
This configuration results in a total of approximately 1000 Markov chain steps, equivalent to $L_\text{HMC}$.
\color{red}
\marginpar{{\footnotesize [R1.1]\par[R3.1]}}%
The warm-up period within $2N$ is set to $12$ and $12\times2$ for uncoded and coded cases, respectively.
Our simulation implementation employs Stan~\cite{stan2023}, a prominent probabilistic programming language which incorporates NUTS.
All remaining Stan parameters use default values, including random initialization of target variables.
\color{black}

\subsubsection{Parameters of Existing Stochastic Methods}
\color{red}
\marginpar{\vspace*{0.3em}{\footnotesize [R1.5]\par[R2.2]\par[R3.4]}}%
Based on the settings specified in Section~\ref{sec:Previous work}, we determine the number of MR such that $\text{MR}_\text{MGS} \times L_\text{MGS}$, $\text{MR}_\text{MHGD} \times L_\text{MHGD}$, and $\text{MR}_\text{Lang} \times T_\text{A} L_\text{Lang}$ are approximately 10,000 to ensure fair computational complexity comparison.
\color{black}

\subsubsection{Well-known Benchmark Methods}
We consider the MMSE as a representative MIMO linear-detection method.

In addition, the MIMO detection efficiency is often measured against single-input single-output (SISO) transmission with additive white Gaussian noise (AWGN), owing to its freedom from cross-antenna interference and channel fading.
The BER performance of the SISO AWGN serves as a crucial reference for evaluating the MIMO detection performance.

\subsection{Computational Complexity}
\color{black}
Although the proposed method for the coded case and annealed Langevin contain an SVD-related term $2N2M \times \min(2N, \allowbreak 2M)$, we ignore it since the iterative term proportional to $(2N)^2$ dominates in our settings.
Consequently, our proposed method for both uncoded and coded cases, MGS, MHGD, and annealed Langevin are on the order of $10000 (2N)^2$, while EP and OAMP-Net2 are on the order of $10 (2N)^3$.
All of these methods demonstrate polynomial-order computational complexities.
\color{black}
The superiority between the two categories varies depending on the operational conditions.
For $N < 1000/2$, which includes our simulation setting of $N = 96$, EP and OAMP-Net2 require less computational cost.
However, even accounting for the SVD-related term, the proposed method, MGS, MHGD, and annealed Langevin achieve lower computational complexity for approximately $N > 1000/2$.
\color{black}

\color{red}
\subsection{HMC Behavior} \label{subsec:MCMC Behavior}
\marginpar{{\footnotesize [R1.2]\par[R1.3]\par[R2.5]\par[R3.6]}}%
Multiple approaches exist to assess the adequacy of MCMC behavior.
Trace plots and autocorrelation plots are first used for visual inspection.
For more precise evaluation, sample statistics such as effective sample size (ESS), potential scale reduction factor $\hat{R}$, and convergence rate $r$ are commonly used.
ESS helps estimate convergence accuracy, with a general guideline of at least 10 and preferably 100 or more per chain for adequate sample size~\cite{BDA3}.
Suppose Markov chains with $I$ steps and $J$ chains.
ESS is defined as ${I \times J}/{(1+2\sum_{\text{Lag}=1}^{T} \widehat{\text{ACF}}_\text{Lag})}$, where $\widehat{\text{ACF}}_\text{Lag}$ is the estimate of the autocorrelation coefficient at Lag averaged across all chains.
$T$ is the first odd positive integer for which $\widehat{\text{ACF}}_{T+1} + \widehat{\text{ACF}}_{T+2}$ is negative.
$\hat{R}$ helps assess the degree of convergence, where values closer to 1 indicate better stability, with a general guideline of $\hat{R} < 1.1$~\cite{BDA3}.
Let $\psi_{i,j}$ be a sample for $u_n$ at the $i$th step and the $j$th chain,
$\hat{R}$ is defined as $\sqrt{{\widehat{\text{Var}}^+(u_n \mid \mathbf{y})}/{W}}$, where 
\begin{equation}
\begin{aligned}
& \widehat{\mathrm{Var}}^{+}(u_n \mid \mathbf{y}) = {(I-1)}/{I} \times W + {1}/{I} \times B, \\[-0.5ex]
& W = \frac{1}{J} \sum_{j=1}^{J} 
\frac{1}{I-1} \sum_{i=1}^{I} (\psi_{i,j} - \bar{\psi}_{j})^2, \, \bar{\psi}_{j} = \frac{1}{I} \sum_{i=1}^{I} \psi_{i,j}, \\[-1.0ex]
& B = \frac{I}{J-1} \sum_{j=1}^{J}(\bar{\psi}_{j} - \bar{\psi}_{})^2, \, \bar{\psi}_{} = \frac{1}{J} \sum_{j=1}^{J} \bar{\psi}_{j}.
\end{aligned}
\end{equation}
$r$ helps estimate convergence efficiency, ranging from 0 to 1, where smaller values indicate faster convergence.
It is defined as the second-largest eigenvalue modulus of the state transition matrix~\cite{Convergencerate}.
While Markov chain characteristics are ultimately reflected in the BER across multiple trials, here we examine detailed behavior in a typical uncoded QPSK result for $\rho = 0$ through visual inspection (one trial of $u_n$) and statistics (100 trials of $\ubt$).
\begin{figure*}[!tb]
\centering
  \begin{minipage}[t]{0.32\linewidth}
    \centering
    \includegraphics[clip,page=1,width=\columnwidth]{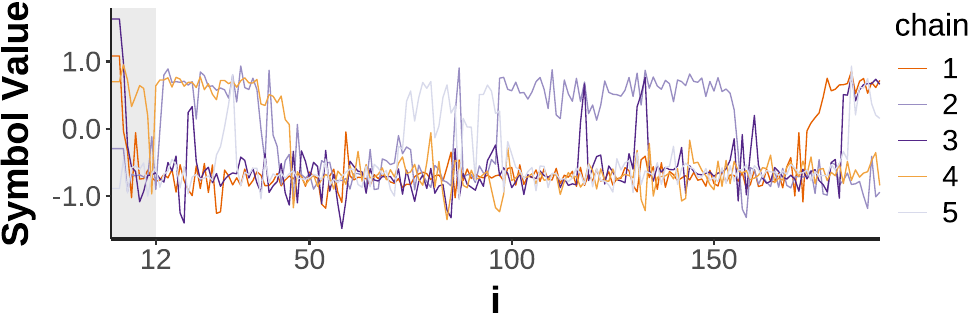}
  \end{minipage}
  \hspace{0.00\linewidth}
  \begin{minipage}[t]{0.32\linewidth}
    \centering
    \includegraphics[clip,page=2,width=\columnwidth]{traceplot_acfplot-crop.pdf}
  \end{minipage}
  \hspace{0.00\linewidth}
  \begin{minipage}[t]{0.32\linewidth}
    \centering
    \includegraphics[clip,page=3,width=\columnwidth]{traceplot_acfplot-crop.pdf}
  \end{minipage}
  \\[0.5em]
  \begin{minipage}[t]{0.32\linewidth}
    \centering
    \includegraphics[clip,page=4,width=\columnwidth]{traceplot_acfplot-crop.pdf}
  \end{minipage}
  \hspace{0.00\linewidth}
  \begin{minipage}[t]{0.32\linewidth}
    \centering
    \includegraphics[clip,page=5,width=\columnwidth]{traceplot_acfplot-crop.pdf}
  \end{minipage}
  \hspace{0.00\linewidth}
  \begin{minipage}[t]{0.32\linewidth}
    \centering
    \includegraphics[clip,page=6,width=\columnwidth]{traceplot_acfplot-crop.pdf}
  \end{minipage}
\caption{Uncoded QPSK ($\rho = 0$): trace plots and autocorrelation plots.}
\label{fig:traceplot_acfplot}
\end{figure*}
\begin{figure}[!tb]
\centering
\includegraphics[width=1.0\linewidth]{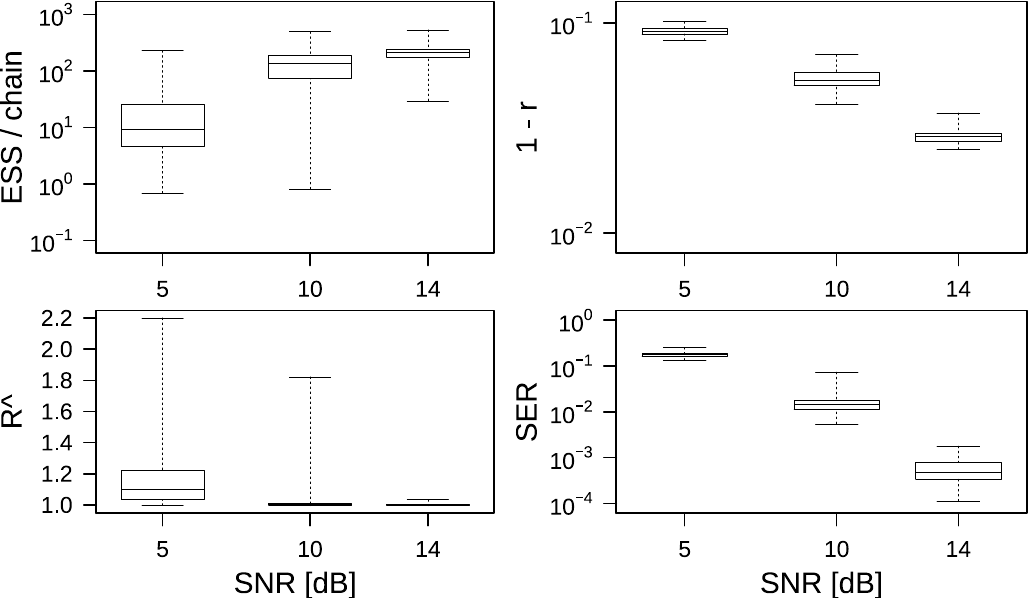}
\caption{Uncoded QPSK ($\rho = 0$): ESS/chain (top left), $\hat{R}$ (bottom left), $1-r$ (top right), and SER (bottom right).}
\label{fig:ESS_R_hat_r_SER}
\end{figure}
\subsubsection{Trace Plot (Fig.~\ref{fig:traceplot_acfplot} top row)}
We set the warm-up period to $12$ for the uncoded case and $12 \times 2$ for the coded case.
The initial severe instability mostly subsides during the warm-up period.
Note that all subsequent analyses in this paper exclude data from the warm-up period.
Following warm-up, we observe active exploration across multiple candidates at low SNR, whereas the sampling focuses on specific candidates as SNR increases.
This occurs because the prior dominates at low SNR while the likelihood dominates at high SNR.

\subsubsection{Autocorrelation Plot (Fig.~\ref{fig:traceplot_acfplot} bottom row)}
The autocorrelation decreases slowly at low SNR, drops rapidly except for chain 4 at moderate SNR, and quickly diminishes to small values for all chains at high SNR.

\subsubsection{ESS and $\hat{R}$ (Fig.~\ref{fig:ESS_R_hat_r_SER} left column)}
Deviations from the guideline are prominent at low SNR but limited at moderate-to-high SNRs.

\subsubsection{$r$ and Symbol Error Rate (SER) (Fig.~\ref{fig:ESS_R_hat_r_SER} right column)}
The convergence rate $r$ of the Markov chain does not necessarily correlate with symbol detection accuracy.
As stated in \cite{Hassibi}, ``there is a trade-off between faster mixing of the Markov chain and faster finding the optimal solution in the stationary distribution.''
For example, high SNR generally induces a slow convergence phenomenon known as Markov chain stalling; however, stalling based on correct symbols is acceptable from a signal detection perspective.
Therefore, we examine both $r$ and SER in this study.
We compute $r$ as follows.
Since we extend the discrete problem to continuous space, we first calculate the responsibility $\gamma^{(k)}_{i,j} = p(s_k \mid \psi_{i, j})$ of continuous sample $\psi_{i, j}$ to discrete symbol $s_k$ using the prior $p_1(\ubt)$.
We then calculate the expected transition counts $C^{(k \to k^\prime)}_j = \sum_i \gamma^{(k)}_{i,j} \gamma^{(k^\prime)}_{i+1, j}$ and average them over $j$ and $n$ to construct the transition matrix.
The second-largest eigenvalue modulus of this transition matrix corresponds to $r$.
The SER is calculated by averaging $\sum_{k^\prime \neq\, \text{correct } k} \gamma^{(k^\prime)}_{i,j}$ over $i$, $j$, and $n$, resulting in an upper-bound estimate compared to hard decision.
Fig.~\ref{fig:ESS_R_hat_r_SER} right column confirms the fundamental MCMC trade-off where convergence slows down as SNR increases while SER decreases.
Although $r$ is high at moderate-to-high SNRs, this poses no practical issue due to the low SER.
Conversely, while $r$ improves somewhat at low SNR, the SER becomes higher.

\subsubsection{Summary}
From a practical perspective, the Markov chain convergence is not necessarily sufficient at low SNR but is deemed satisfactory at moderate-to-high SNRs.
As will be verified later, this result is consistent with the BER performance.

\begin{figure}[!tb]
\centering
\includegraphics[clip,page=1,width=0.7\linewidth]{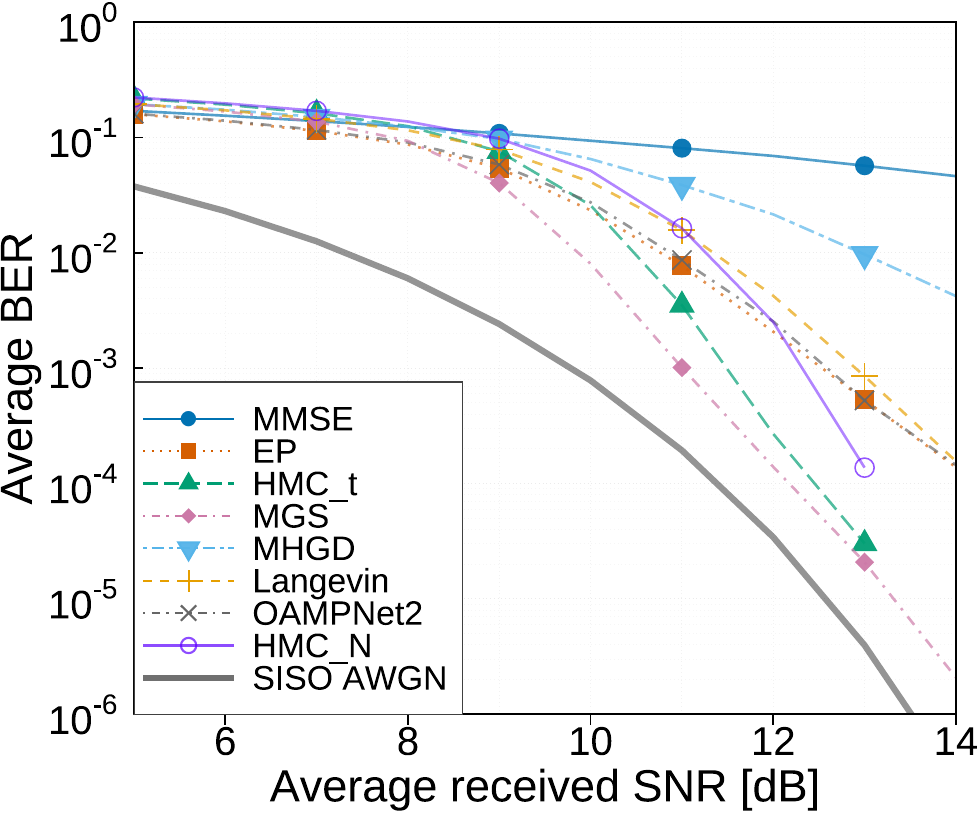}
\caption{Uncoded QPSK ($\rho = 0.5$): average BER versus average received SNR.}
\label{fig:normal_vs_t}
\end{figure}

\subsection{Prior Distribution Choice: Mixture of Normal Distributions vs. $t$-Distributions (Fig.~\ref{fig:normal_vs_t})}
\marginpar{\footnotesize [R1.4]\par[R3.3]}%
We verify the prior distribution choice through the BER performance of uncoded QPSK.
The setting of the normal mixture prior follows~\cite{Globecom}.
When comparing the normal mixture prior (HMC\_N) and $t$-mixture prior (HMC\_t) for $\rho=0.5$, where signal detection is challenging and differences are apparent, HMC\_t clearly demonstrates superior performance.
Note that for higher-order modulations beyond QPSK, the benefit of improved exploration efficiency with the $t$-mixture prior relatively diminishes due to increased constellation points. Nevertheless, it never becomes counterproductive compared to the normal mixture prior~\cite{Globecom}.
\color{black}

\begin{figure*}[!tb]
\centering
  \begin{minipage}[t]{0.33\linewidth}
    \centering
    \includegraphics[clip,page=1,width=\columnwidth]{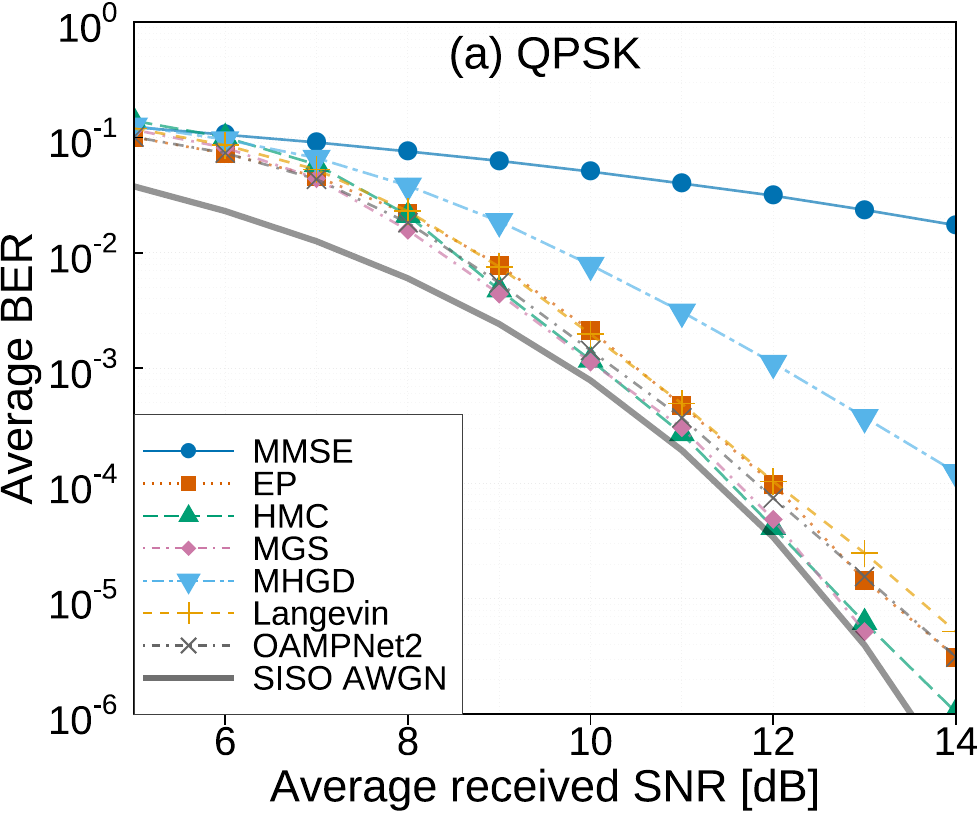}
  \end{minipage}
  \hspace{-0.01\linewidth}
  \begin{minipage}[t]{0.33\linewidth}
    \centering
    \includegraphics[clip,page=2,width=\columnwidth]{BER-crop.pdf}
  \end{minipage}
 \hspace{-0.01\linewidth}
  \begin{minipage}[t]{0.33\linewidth}
    \centering
    \includegraphics[clip,page=3,width=\columnwidth]{BER-crop.pdf}
  \end{minipage}
\caption{Uncoded case ($\rho = 0$): average BER versus average received SNR.}
\label{fig:tVSE}
\end{figure*}

\subsection{Uncoded BER Performance for $\rho=0$: Proposed Method vs.~Linear MMSE Method (Fig.~\ref{fig:tVSE} (a) Through (c))}
The proposed method significantly outperforms the MMSE at moderate-to-high SNRs for all modulation orders.
This confirms the advantage of our nonlinear processing method over linear methods.
However, the proposed method performs equally or slightly worse at low SNRs for all modulation orders.
The factors causing performance degradation at low SNR and their countermeasures are discussed in Sections~\ref{subsec:first_decoding} and \ref{subsec:second_decoding}.

\subsection{Uncoded BER Performance for $\rho=0$: Proposed Method vs.~Existing Stochastic Methods}

\subsubsection{Overall Trend (Fig.~\ref{fig:tVSE} (a) Through (c))}
Increasing the modulation order results in greater performance divergence from the SISO AWGN benchmarks.
Higher modulation orders expand the signal constellations, reducing inter-point distances and complicating the signal estimation processes.

\subsubsection{Comparison for QPSK (Fig.~\ref{fig:tVSE} (a))}
The performances of all the examined stochastic methods, excluding the MHGD, are closely aligned with that of the SISO AWGN, showing slight variations.
The simplicity of signal detection resulting from the sparse transmission points and wide inter-point spacing for the QPSK contributes to this similarity in performance.
\color{red}
\marginpar{{\footnotesize [R2.2]}}%
The inferior performance of MHGD can be attributed to the quantization to the nearest constellation point at each Markov chain step, which violates the exact Markov chain property. This quantization becomes problematic for large $N$ as in our setting, suffering from combinatorial explosion inherent in discrete optimization.
\color{black}
Under low SNR conditions, our proposed method exhibits slightly inferior performance compared to the existing stochastic methods.
Sections~\ref{subsec:first_decoding} and \ref{subsec:second_decoding} explore low SNR performance decline factors and remedies.
At moderate-to-high SNRs, our proposed method surpasses the three methods (EP, annealed Langevin, and OAMP-Net2) and almost matches the MGS in terms of performance.
At $10^{-3}$ BER, our proposed method achieves SNR gains of approximately 0 dB over MGS, 0.4 dB over EP, 2.0 dB over MHGD, 0.4 dB over annealed Langevin, and 0.2 dB over OAMP-Net2.

\subsubsection{Comparison for 16QAM (Fig.~\ref{fig:tVSE} (b))}
Unlike the case in the QPSK scenario, our proposed method achieves the best detection performance at moderate-to-high SNRs.
The performance gap between our proposed method and the two methods (EP and OAMP-Net2) widens at moderate-to-high SNRs compared to the case in the QPSK scenarios.
\color{red}
\marginpar{{\footnotesize [R3.8]}}%
This is because these methods incorporate the concept of approximation using a unimodal distribution, with details to be discussed in the 64QAM scenario.
\color{black}
At $10^{-3}$ BER, our proposed method achieves SNR gains of 0.4 dB over MGS, 1.4 dB over EP, 2.8 dB over MHGD, 0.4 dB over annealed Langevin, and 1.6 dB over OAMP-Net2.

\subsubsection{Comparison for 64QAM (Fig.~\ref{fig:tVSE} (c))} 
Unlike the case in the QPSK and 16QAM scenarios,
\color{red}
\marginpar{{\footnotesize [R1.5]\par[R3.4]}}%
MGS shows significant degradation.
This is because our setting for 64QAM uses $\text{MR}_\text{MGS} = 2$, which is much lower than $\text{MR}_\text{MGS} = 50$ suggested in~\cite{MGS}, to align computational complexity.
\color{black}
Our proposed method excels at moderate-to-high SNRs, as is the case in the 16QAM scenario.
Furthermore, in the 64QAM scenario, our performance advantage over existing stochastic methods becomes more pronounced.
At $10^{-3}$ BER, our proposed method achieves SNR gains of more than 4.5 dB over MGS, 4.1 dB over EP, 3.2 dB over MHGD, 1.5 dB over annealed Langevin, and more than 4.5 dB over OAMP-Net2.

\begin{figure}[!tb]
\centering
\includegraphics[draft=false,clip,width=0.9\linewidth]{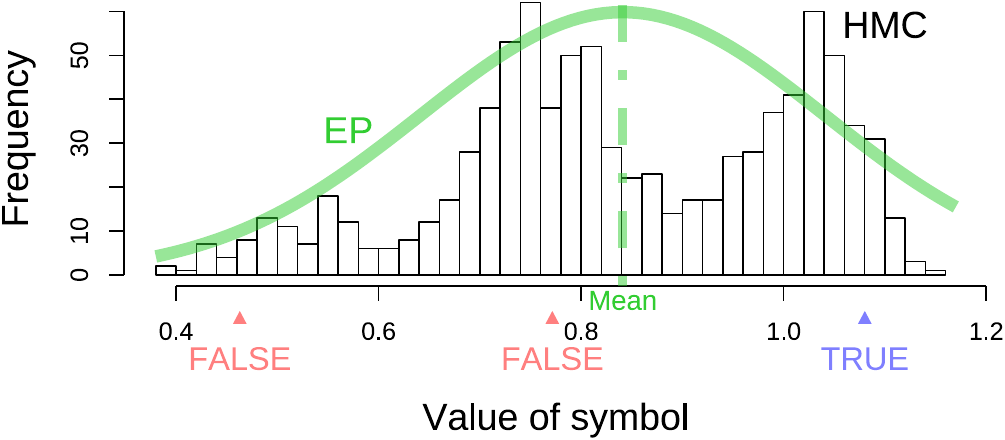}
\caption{Uncoded 64QAM ($\rho=0$ and SNR = 25 dB): histogram of the marginal posterior distribution for a certain symbol.}
\label{fig:why64QAMsperior}
\end{figure}
\begin{figure*}[!t]
\centering
  \begin{minipage}[t]{0.33\linewidth}
    \centering
    \includegraphics[draft=false,clip,page=1,width=\columnwidth]{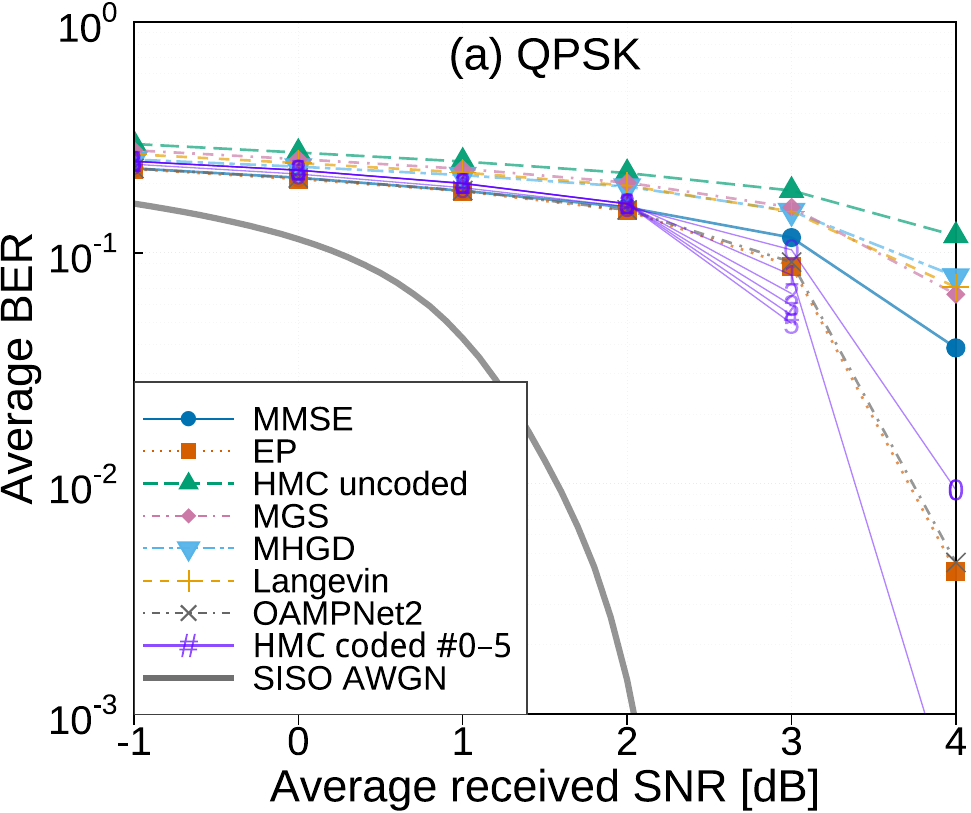}
  \end{minipage}
  \hspace{-0.01\linewidth}
  \begin{minipage}[t]{0.33\linewidth}
    \centering
    \includegraphics[draft=false,clip,page=2,width=\columnwidth]{coded_BER-crop.pdf}
  \end{minipage}
 \hspace{-0.01\linewidth}
  \begin{minipage}[t]{0.33\linewidth}
    \centering
    \includegraphics[draft=false,clip,page=3,width=\columnwidth]{coded_BER-crop.pdf}
  \end{minipage}
\caption{Coded case ($\rho = 0$): average BER versus average received SNR.}
\label{fig:codedBER}
\end{figure*}
Here, based on a specific example, we explain why the performance advantage of the proposed method relatively grows with the modulation order.
This trend is related to the ability of the method to express multimodality in the posterior distribution.
For example, the EP assumes a unimodal multivariate normal distribution for the approximate posterior distribution.
In contrast, our simulation-based method can capture the posterior distribution as it is, thus expressing multimodality.
Fig.~\ref{fig:why64QAMsperior} shows a histogram of the marginal posterior distribution for a certain symbol where the BER performance differs.
In difficult estimation scenarios, the posterior distribution becomes multimodal as multiple possibilities for the transmission symbols remain plausible, as shown in this example.
Forcibly approximating such a distribution with a unimodal distribution can lead to erroneous symbol estimations.
That is, the unimodal peak may shift incorrectly due to the influence of the wrong symbol possibilities.
The proposed method can select the result with the maximum likelihood among multiple modes, enabling correct symbol estimation and improving the detection performance.
Such a situation becomes more likely as the number of candidate transmission symbols increases.
Thus, the effectiveness of the proposed method increases with the modulation order.
\color{red}
\marginpar{{\footnotesize [R1.5]\par[R2.2]\par[R3.4]\par[R3.8]}}%
Likewise, OAMP-Net2 does not directly consider multimodality in the posterior distribution, which likely explains its similar degradation trends to those of EP.
In contrast, MGS, MHGD, and annealed Langevin can represent multimodal posterior distributions but show inferior performance compared to the proposed method.
This is presumably because performance differences among exploration algorithms become evident in challenging high-order modulations.
As mentioned earlier, MHGD suffers from quantization at each MCMC step, while MGS is hampered by insufficient MR iterations.
Furthermore, annealed Langevin solves the Hamiltonian equations for only one step, unlike HMC, limiting exploration diversity. This may increase the risk of local optima as modulation order rises.
\color{black}

\subsection{Uncoded BER Performance for $\rho=0$: Proposed Method vs.~SISO AWGN (Fig.~\ref{fig:tVSE} (a) Through (c))} 
Our proposed method exhibits near-optimal performance for signal detection in the uncoded case.
At $10^{-3}$ BER, the SNR degradation for our method remains below 0.3, 1.3, and 2.9 dB in the QPSK, 16QAM, and 64QAM scenarios, respectively.
This remarkable efficacy stems from a well-tuned prior distribution configuration coupled with an efficient search algorithm.

\color{red}
\subsection{Effects Under Realistic Operating Conditions}

\subsubsection{Spatial Correlation (Fig.~\ref{fig:normal_vs_t})}
\marginpar{{\footnotesize [R2.7]\par[R3.2]}}%
We investigate this effect by comparing the proposed method with other stochastic approaches using uncoded QPSK results for $\rho = 0.5$.
While all methods degrade compared to the $\rho=0$ case, the proposed method (HMC\_t) achieves the second-best performance after MGS at moderate SNR and approaches the best-performing MGS at high SNR.
Note that the proposed method's poor performance at low SNR follows the same trend as in the $\rho=0$ case.

\subsubsection{Imperfect Channel Estimation}
\marginpar{{\footnotesize [R2.3]\par[R3.2]}}%
For simplicity, we examine the fundamental effects for the case of $\rho=0$.
Let $\Hbt^\prime$ denote the estimated channel via MMSE using orthogonal pilots of length $2N$; then, the channel estimation error is $\Delta \Hbt = \Hbt - \Hbt^\prime$.
Since $\Hbt^\prime$ is obtained through linear operations on received orthogonal pilots with $\rho=0$, it consists of i.i.d. zero-mean normal random variables, as does $\Delta \Hbt$.
From~\eqref{eq:system}, we have $\ybt = \Hbt^\prime \ubt + \wbt^\prime$, where $\wbt^\prime = \wbt + \Delta \Hbt \ubt$.
Since $\Hbt^\prime$ is obtained via MMSE, it can be treated as known.
Consequently, channel estimation errors modify only the noise term in the system model~\eqref{eq:system}.
Under the square constellation assumption where $\text{E}[\ubt] = \zero$, both $\Delta \Hbt \ubt$ and $\wbt^\prime$ follow zero-mean normal distributions.
The variance of $\Delta \Hbt \ubt$ is simply $1/\{1/\sigma_w^2 + 1/(2N P_t)\} \Ibt$ under $\rho=0$ with orthogonal pilots~\cite{ImperfectCSI}.
Therefore, the variance of $\wbt^\prime$ is $\sigma_w^2 \Ibt + 1/\{1/\sigma_w^2 + 1/(2N P_t)\} \Ibt = [1 + 1/\{1+ \sigma_w^2 / (2N P_t)\}] \sigma_w^2 \Ibt$.
Ultimately, for $\rho=0$, the effect of channel estimation errors with orthogonal pilots is equivalent to SNR degradation in~\eqref{eq:system}.
Since channel estimation error effects vanish as $\sigma_w^2 \to \infty$, higher SNR regions suffer more severe impacts.
For example, the limiting case of $\sigma_w^2 \to 0$ yields a 3 dB penalty.
\color{black}

\subsection{Coded BER Performance for $\rho=0$: Initial Detection and Decoding Without LLR Feedback (Fig.~\ref{fig:codedBER} (a) Through (c))}\label{subsec:first_decoding}
For reference, we first present the results obtained by simply applying error-correction codes to the previous uncoded method (``HMC uncoded'' in Fig.~\ref{fig:codedBER}).
Based on the results, the proposed method exhibits the worst performance.
This is due to its poor performance at low SNRs in the uncoded case.
Therefore, we consider improving the performance of the proposed method at low SNRs.
The relatively good performances of the three methods (MMSE, EP, and OAMP-Net2) provide insights for this improvement.

The MMSE results suggest that ridge regularization, which mitigates noise enhancement, is effective in noisy situations.
Therefore, we introduce an additional prior distribution $p_2(\ubt)$.

The EP and OAMP-Net2 estimate the transmitted symbols by assuming unimodal approximation.
This feature is a disadvantage in the ambiguous estimation scenarios at moderate-to-high SNRs, as shown in Fig.~\ref{fig:why64QAMsperior}.
However, the average estimates may be advantageous at low SNRs.
Consequently, we change the point estimator of the posterior distribution to the marginal posterior-based point estimator.

\color{red}
\marginpar{{\footnotesize [R2.8]\par[R3.7]}}%
Considering the above, we use the settings shown in the initial phase column of Table~\ref{tbl:coded_summary}.
\color{black}
The BER performances labeled ``HMC coded \#0'' in Fig.~\ref{fig:codedBER} display the results under these conditions.
This approach achieves performance better than that of MMSE and closer to that of EP and OAMP-Net2 in the waterfall region for all modulation orders.

\subsection{Coded BER Performance for $\rho=0$: Subsequent Joint Detection and Decoding With LLR Feedback (Fig.~\ref{fig:codedBER} (a) Through (c))}\label{subsec:second_decoding}
Here, we consider further improving the coded performance of the proposed method using the LLR feedback.
\color{red}
\marginpar{{\footnotesize [R2.8]\par[R3.7]}}%
Specifically, we configure the likelihood as the horseshoe density, maintain the prior distribution with ridge regularization, and employ the joint posterior-based point estimator for the posterior distribution.
Based on this configuration detailed in the subsequent phases column of Table~\ref{tbl:coded_summary}, we present the BER performances  (``HMC coded \#1--5'' in Fig.~\ref{fig:codedBER}).
\color{black}
In the waterfall region, the performance of the proposed method improves as the number of iterations increases.
The method ultimately outperforms EP and OAMP-Net2, showing the best results among the compared methods.
The improvement in the estimation accuracy with the number of iterations is more pronounced at higher modulation orders.
This trend matches the uncoded case, believed to stem from the posterior distribution's multimodality and HMC's strong exploration.

\begin{figure}[!tb]
\centering
\includegraphics[clip,width=0.7\linewidth]{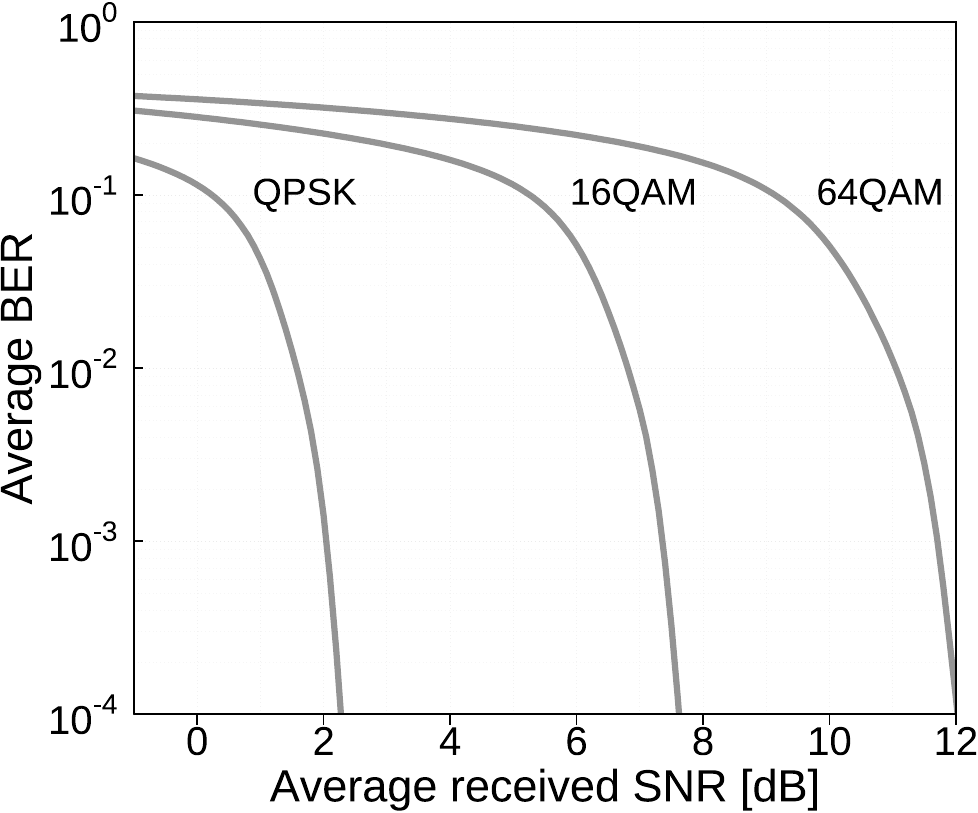}
\caption{Coded BER performances of SISO AWGN.}
\label{fig:siso_awgn_ber}
\end{figure}

Finally, we compare the performance with the SISO AWGN.
In Fig.~\ref{fig:codedBER}, the SISO AWGN plot for the 64QAM scenario is out of range.
Thus, for proper representation, Fig.~\ref{fig:siso_awgn_ber} shows the SISO AWGN performances for all modulation orders.
For all modulation orders, the final performance of the proposed method approaches that of the SISO AWGN.
For QPSK at $10^{-1}$ BER, the SNR degradation of our method remains below 2.2 dB.
For 16QAM and 64QAM at $10^{-2}$ BER, it remains below 1.9 and 2.2 dB (using the 4th iteration for 64QAM), respectively.
Therefore, even for coded signal detection, our proposed method exhibits near-optimal performance.

\begin{figure}[!tb]
\centering
  \begin{minipage}[t]{0.49\linewidth}
    \centering
    \includegraphics[clip,page=1,width=\linewidth]{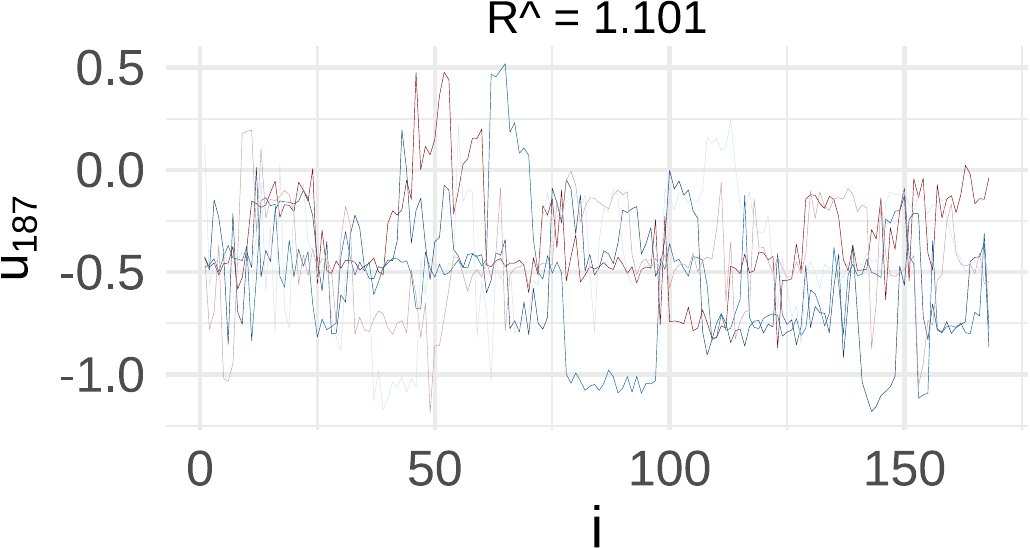}
  \end{minipage}
  \hfill
  \begin{minipage}[t]{0.49\linewidth}
    \centering
    \includegraphics[clip,page=2,width=\linewidth]{traceplot_HS-crop.pdf}
  \end{minipage}
\\[0.5em]
  \begin{minipage}[t]{0.49\linewidth}
    \centering
    \includegraphics[clip,page=3,width=\linewidth]{traceplot_HS-crop.pdf}
  \end{minipage}
  \hfill
  \begin{minipage}[t]{0.49\linewidth}
    \centering
    \includegraphics[clip,page=4,width=\linewidth]{traceplot_HS-crop.pdf}
  \end{minipage}
\\[-0.3em]
  \begin{minipage}[t]{0.45\linewidth}
    \centering
    \includegraphics[clip,page=5,width=\linewidth]{traceplot_HS-crop.pdf}
  \end{minipage}
\caption{Coded 64QAM ($\rho=0$ and SNR = 14 dB): initial phase (top left), first subsequent phase with normal likelihood (top right), first subsequent phase with horseshoe likelihood (bottom left), and $\lambda_{187}$ in horseshoe likelihood (bottom right).}
\label{fig:HS}
\end{figure}

\color{red}
\subsection{Effectiveness of the Horseshoe Likelihood (Fig.~\ref{fig:HS})}
\marginpar{{\footnotesize [R1.1]\par[R2.1]\par[R3.3]}}%
We examine one typical trial of $u_{187}$ (true value: $-0.463$) through the trace plot and $\hat{R}$ for coded 64QAM ($\rho=0$ and SNR = 14 dB).
The initial phase (Fig.~\ref{fig:HS} top left) shows exploration of multiple symbol positions across positive and negative regions, yielding an $\hat{R}$ value near the high threshold.
This occurs because the Markov chain does not necessarily converge at low SNR, as explained in Section~\ref{subsec:MCMC Behavior}.
We then examine the first subsequent phase following LLR feedback from the LDPC decoder.
When we naively use a normal likelihood (Fig.~\ref{fig:HS} top right), LLR feedback improves Markov chain convergence and focuses the exploration on the correct negative region. However, it continues to explore multiple symbol positions, and $\hat{R}$ remains slightly elevated.
If we could increase the temperature parameter in the likelihood, the broadened likelihood and resulting smoother posterior would help improve convergence, but its optimization remains difficult~\cite{Hassibi}.
To address this challenge, parallel tempering with multiple temperature parameters has been proposed~\cite{BDA3}, but it increases computational burden proportionally.
Treating the temperature parameter as a random variable, as in our approach, enables dynamic tempering (similar to adaptive parallel tempering) that efficiently addresses convergence issues only when needed, improving exploration efficiency.
With horseshoe likelihood (Fig.~\ref{fig:HS} bottom left), exploration concentrates around the true value with improved $\hat{R}$.
Furthermore, the trace plot and boxplot of temperature coefficient $\lambda_{187}$ (Fig.~\ref{fig:HS} bottom right) reveal occasional extreme values.
This demonstrates selective improvement of Markov chain convergence.
The heavy-tailed Cauchy distribution suits this application as a prior by accommodating occasional extreme values.

\subsection{Limitations and Extensions of the Proposed Method}
\marginpar{{\footnotesize [R2.6]\par[R3.1]}}%
While we have examined the characteristics for $N = M = 96$ in our simulations, we now discuss extensions for larger antenna configurations.
First, the optimal parameter values obtained through a preliminary search may vary with the number of antennas.
To overcome this issue, deep unfolding can be applied.
The necessary MCMC iterations and number of chains may increase more rapidly than anticipated; however, parallel chains can be efficiently processed using contemporary multicore processor architectures.
Moreover, in typical massive MIMO applications such as multiuser MIMO downlink, an overloaded scenario with $N > M$ is expected.
In this case, rank deficiency of $\Hbt$ severely degrades BER performance, but ridge regularization can mitigate this issue, suggesting the application of $p_2(\ubt)$ even in the uncoded case.

\marginpar{{\footnotesize [R2.7]\par[R3.2]}}%
Although this study assumed normal noise, this assumption may not always hold in practice.
For example, in the presence of inter-cell interference, the aggregate interference can be treated as non-normal noise~\cite{nonGaussiannoise}.
Nevertheless, when the noise follows typical continuous distributions such as gamma or $\alpha$-stable distributions, our method can be easily extended by redefining the likelihood function.
\color{black}

\section{Conclusion} \label{sec:Conclusions}
Employing HMC, $t$-distribution mixture priors, and horseshoe likelihood, we have developed a signal-detection method capable of delivering near-optimal performance with polynomial complexity.
For higher-order modulations, our method significantly outperforms conventional methods in terms of detection performance.
Higher-order modulations are poised to play a pivotal role in next-generation wireless networks.
Wireless data traffic is expected to surge, driven by multisensory interactions surpassing conventional voice and video communication.
Consequently, our proposed approach holds significant potential for advancing future wireless technologies.

This research, however, has a limitation that warrants consideration.
\color{red}
\marginpar{{\footnotesize [R2.6]\par[R2.7]\par[R3.1]}}%
Specifically, the discussion is limited to 96 transmit and receive antennas and normal noise. Although detailed investigation beyond these constraints remains future work, this does not diminish the value of this study.
\color{black}

Mathematically, our findings suggest that continuous extensions of discrete problems may yield superior solutions.
Our approach could provide valuable insights for continuous relaxation in combinatorial optimization, potentially benefiting classic problems such as the traveling salesman and nurse scheduling problems.

\appendices

\section*{Acknowledgment}
The authors express their sincere gratitude to Kazushi Matsumura, Hiroki Asumi, and Yukiko Kasuga for their invaluable contributions to this research.
They were undergraduate and graduate students at Hokkaido University during the study.
\color{red}
This work was supported by JSPS KAKENHI Grant Number 25K07741.
\color{black}



\end{document}